\documentclass[prl,reprint,twocolumn,aps,showpacs,floatfix]{revtex4-1}
\usepackage{amsmath}
\usepackage{amssymb}
\usepackage{graphicx}
\usepackage{siunitx}
\usepackage{braket}
\usepackage[colorlinks=true,linkcolor=blue,citecolor=blue,urlcolor=blue]{hyperref}
\usepackage{txfonts}
\usepackage{xr}
\usepackage{color}
\definecolor{RED}{rgb}{1,0,0}
 \definecolor{BLUE}{rgb}{0,0,1}
\begin{document}
\title{Resolution of the exponent puzzle for the Anderson transition in doped semiconductors}
\author{Edoardo G. Carnio}
\email{e.carnio@warwick.ac.uk}
\affiliation{Department of Physics, University of Warwick, Coventry, CV4 7AL,
	United Kingdom}

\author{Nicholas D. M. Hine}
\affiliation{Department of Physics, University of Warwick, Coventry, CV4 7AL,
	United Kingdom}

\author{Rudolf A. R\"omer}
\affiliation{Department of Physics, University of Warwick, Coventry, CV4 7AL,
	United Kingdom}
%\affiliation{Centre for Scientific Computing, University of Warwick, Coventry, CV4 7AL, United Kingdom}


\begin{abstract}
The Anderson metal-insulator transition (MIT) is central to our understanding of the quantum mechanical nature of disordered materials. Despite extensive efforts by theory and experiment, there is still no agreement on the value of the critical exponent $\nu$ describing the universality of the transition --- the so-called ``exponent puzzle''.
In this work, going beyond the standard Anderson model, we employ \textit{ab initio} methods to study the MIT in a realistic model of a doped semiconductor. 
We use linear-scaling DFT to simulate prototypes of sulfur-doped silicon (Si:S). From these we build larger tight-binding models close to the critical concentration of the MIT. 
When the dopant concentration is increased, an impurity band forms and eventually delocalizes.
We characterize the MIT via multifractal finite-size scaling, obtaining the phase diagram and estimates of $\nu$. 
Our results suggest an explanation of the long-standing exponent puzzle, which we link to the hybridization of conduction and impurity bands.
\end{abstract}

\maketitle
%%%%%%%%%%%%%%%%%%%%%%%%%%%%%%%%%%%%%%%%%%%%%%%%%%%%%%%%%%%%

The Anderson metal-insulator transition (MIT) is the paradigmatic quantum phase transition, resulting from spatial localization of the electronic wave function due to increasing disorder \cite{Anderson1958a}. As for any such transition, universal critical exponents capture its underlying fundamental symmetries. This universality allows to disregard microscopic detail and the Anderson MIT is expected to share a single set of exponents.
The last decade has witnessed many ground-breaking experiments designed to observe Anderson localization \emph{directly}: with light \cite{Wiersma1997,Scheffold1999,Johnson2003,Storzer2006,VanderBeek2012,Sperling2012,Scheffold2013,Wiersma2013,Sperling2016,Skipetrov2016}, photonic crystals \cite{Schwartz2007,Wiersma2013}, ultrasound \cite{Hu2008,Faez2009}, matter waves \cite{Billy2008}, Bose-Einstein condensates \cite{Roati2008} and ultracold matter \cite{Kondov2011,Jendrzejewski2012}. The mobility edge \cite{Mott1966}, separating extended from localized states, was only measured for the first time in 2015 \cite{Semeghini2015}. The hallmark of these experiments is the tunability of the experimental parameters and the ability to study systems where many-body interactions are absent or can be neglected. 
Under such controlled conditions, the observed exponential wave function decay, the existence of mobility edges and the critical properties of the transition \cite{Chabe2008,Lopez2012} are in excellent agreement with the non-interacting Anderson model \cite{Anderson1958a}. Furthermore, scaling at the transition \cite{Abrahams1979} leads to high-precision estimates of the \emph{universal} critical exponent $\nu$ from transport simulations ($\nu=1.57(1.55,1.59)$ \cite{Slevin1999}) and wave function statistics ($\nu = 1.590(1.579,1.602)$ \cite{Rodriguez2011}). 

Anderson's original challenge was to describe localization in doped semiconductors. For these ubiquitous materials, the existence of the MIT was confirmed \emph{indirectly} by measuring the scaling of the conductance $\sigma\sim ( n - n_\text{c} )^{\nu}$ when increasing the dopant concentration $n$ beyond its critical value $n_\text{c}$. 
However, a puzzling discrepancy remains: a careful analysis by Itoh et al.\ \cite{ItoWOH04} highlights that the value of $\nu$ can change significantly with the control of dopant concentration around the transition point, the homogeneity of the doping, and the purity of the sample.
Following Stupp et al.\ \cite{Stupp1993}, they suggest that the intrinsic behaviour of an uncompensated semiconductor gives $\nu\approx 0.5$ \cite{Thomas1983}, while any degree of compensation results in $\nu \approx 1$ \cite{Waffenschmidt1999}. Evidently, these values disagree with the aforementioned theoretical and experimental studies. The inability to characterize the Anderson transition in terms of a single, universal value for $\nu$ is known as the ``exponent puzzle''  \cite{Thomas1985,Stupp1993}.

Most theoretical models that have been applied to this problem lack the ability to capture the full complexity of a semiconductor. 
The Anderson model, for example, ignores the detail of the crystal lattice and the electronic structure, and also simplifies the physics by ignoring many-body interactions and interactions between dopant and host material.
These factors are known to change the universal behavior \cite{Izrailev1999,PetS13} and the value of $\nu$, as shown in studies on correlated disorder \cite{Ndawana005,Croy2012,Croy2012a} and hydrogenic impurities in an effective medium, where $\nu \approx 1.3 $ \cite{Harashima2012,Harashima2014}. 
Here we propose a fundamental shift from studying localization using highly-simplified tight-binding Anderson models, to atomistically correct \emph{ab initio} simulations \cite{Neugebauer2013,Jain2016} of a doped semiconductor.
We illustrate the power of our approach for sulfur-doped silicon, Si:S, where the MIT occurs for concentrations in the range $1.8$--$\SI{4.3E20}{cm^{-3}}$ \cite{Winkler2011}. We model the donor distribution in Si:S by randomly placing the impurities in the lattice \cite{Winkler2009}.
While we concentrate on Si:S here, our method is straightforwardly applicable to other types of impurities (Si:P; Si:As; Ge:Sb), hole doping (Si:B) and co-doping (Si:P,B; Ge:Ga,As). 

\begin{figure}[bt]
	\includegraphics[width=\columnwidth]{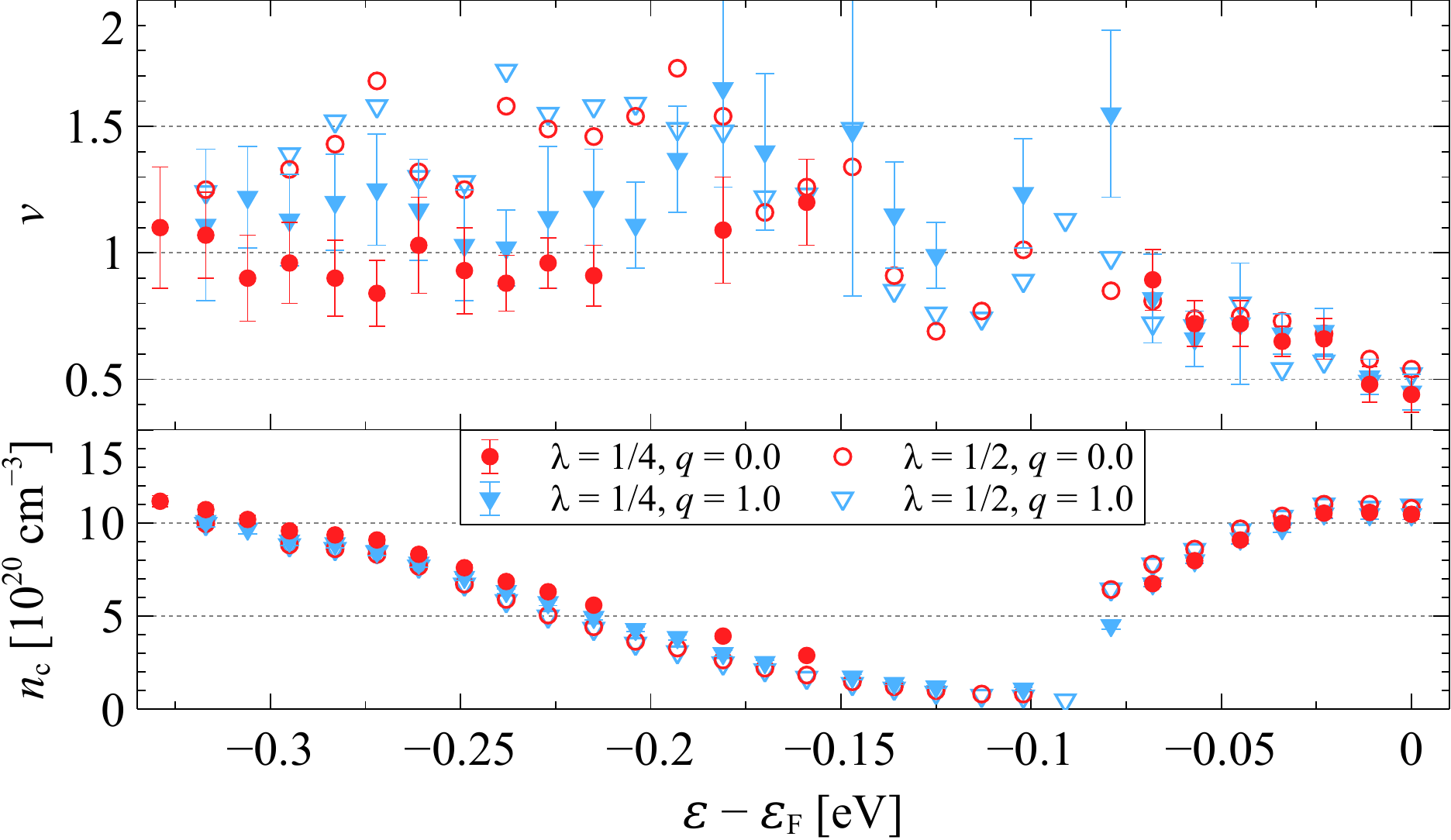} 
	\caption{Critical concentrations $n_\text{c}$ and exponents $\nu$ as a function of the energy $\varepsilon$ from the Fermi level $\varepsilon_\text{F}$, for $q=0$ (red circles) and $q=1$ (blue triangles). Full and open symbols denote, respectively, the results for $\lambda = 1/4$ and $\lambda = 1/2$ coarse-grainings. The error bars, shown only for $\lambda=1/4$ and if larger than symbol size, represent the 95\% confidence level on the fit parameters. The error bars for $\lambda=1/2$ are of the same order of magnitude as for $\lambda=1/4$ and are omitted for clarity.}
	\label{fig:PHDIAG}
\end{figure}

With this approach we observe the formation of the impurity band (IB), upon increasing $n$, and its eventual merger with the conduction band (CB).
States in the IB become delocalized, as measured directly via multifractal statistics of wave functions \cite{Janssen1994}, and we observe and characterize the MIT. 
In Fig.\ \ref{fig:PHDIAG} we plot how  $n_\text{c}$ and $\nu$ vary for energies $\varepsilon$ in the IB below the Fermi energy $\varepsilon_\text{F}$.
%%%%%%%%%%%%%%%%%%%%%%%%%%%%%%%%%%%%%%%%%%%%%%%%%%%%%%%%%%%%
%%%%%%%%%%%%%%%%%%%%%%%%%%%%%%%%%%%%%%%%%%%%%%%%%%%%%%%%%%%%
For $\varepsilon \sim \varepsilon_\text{F}$, the values are $\nu \sim 0.5$, while deeper in the IB the exponents increase to about $\nu \sim 1$, reaching values around $1.5$. As we will show below, our simulations of an uncompensated semiconductor suggest that the reduction in $\nu$ at $\varepsilon_\text{F}$ is due to the hybridization of IB and CB. Deep in the IB the physics of the Anderson transition reemerges with $\nu$ reaching the range of its proposed universal value \cite{Slevin1999,Harashima2014,Garcia-Garcia2008}. 
Experiments can readily access these higher values by moving  $\varepsilon_\text{F}$ via compensation \cite{ItoWOH04} --- intentional or otherwise.

%%%%%%%%%%%%%%%%%%%%%%%%%%%%%%%%%%%%%%%%%%%%%%%%%%%%%%%%%%%%
\begin{figure*}[tb]
	\includegraphics[width=\textwidth]{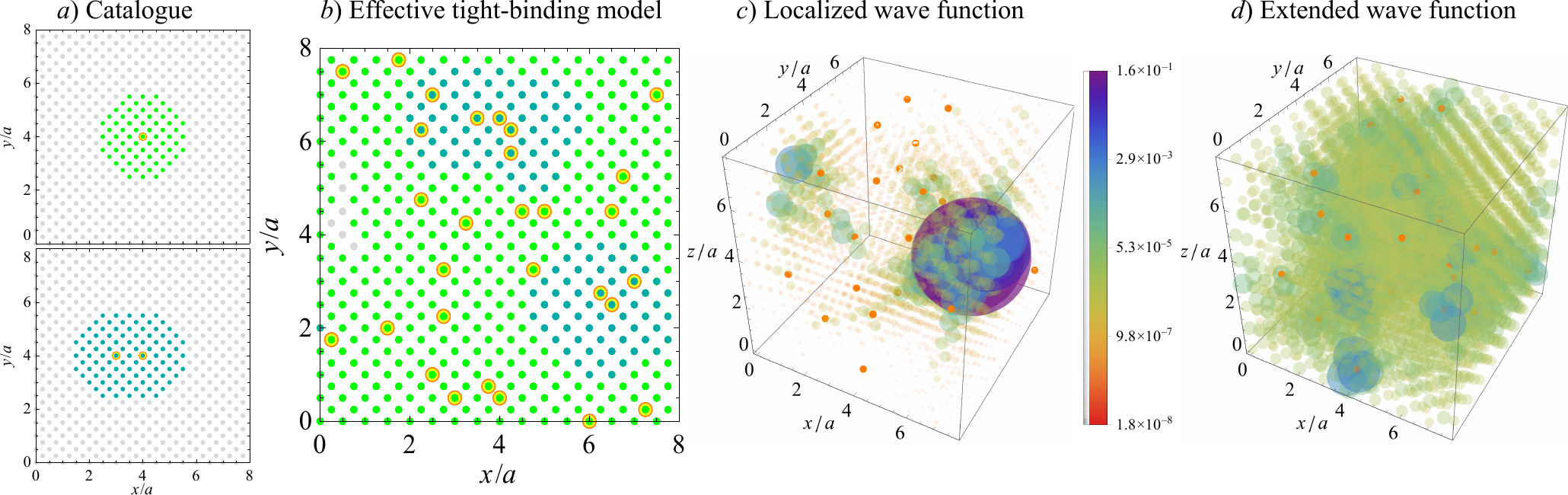} 
	\caption{Schematic description of the work-flow. Part (\textit{a}) represents the catalog of prototypes. For clarity we show a projection on the $ xy $ plane and distances in units of $ a $, the Si lattice parameter. The upper plot depicts one impurity (yellow with orange border) and the neighboring Si atoms (green); the lower plot denotes two impurities at distance $a$ and their Si neighbors (dark green). Gray sites indicate Si atoms unaffected by the impurity potential. In (\textit{b}) we indicate how we build an effective tight-binding model of $ 4096 $ atoms with $ 29 $ impurities. The color code is the same as in (\textit{a}) and indicates which catalog is used. Due to the projection on the $ xy $ plane some impurities appear closer than they are. Their 3D distribution is shown in (\textit{c}) and (\textit{d}), where we plot (\textit{c}) a localized state deep in the IB and (\textit{d}) an extended state above $\varepsilon_\text{F}$.
We represent the 90\%  largest wave function values with spheres of volume proportional to $ |\psi|^2 $. Opacity and color are proportional to $ - \log_L |\psi|^2$, with $ L=16 $ here, so that lower (higher) values are in red transparent (violet solid). The box size is as in (\textit{b}).}
	\label{fig:scheme}
\end{figure*}
%%%%%%%%%%%%%%%%%%%%%%%%%%%%%%%%%%%%%%%%%%%%%%%%%%%%%%%%%%%%

Density functional theory (DFT) calculations are now the method of choice for \emph{ab initio} solid state materials characterization \cite{Neugebauer2013} and discovery \cite{Jain2016}. 
With the choice of Si:S, we can observe the transition in systems of up to $11 \times 11 \times 11$ unit cells, i.e., $10648$ atoms.
These large system sizes can in principle be reached by \emph{linear-scaling} DFT \cite{Hine2009}, but despite this, the necessity to average over many hundreds of disorder realizations, makes repeated DFT calculations impractical for our purposes %
\footnote{A single fixed geometry DFT simulation uses $1152$ cores for about $12$ h on the UK ARCHER HPC facility.}.
We therefore devise a hybrid approach: linear-scaling DFT calculations are performed, using the ONETEP code \cite{Skylaris2005}, on prototype systems of $8 \times 8 \times 8$ diamond-cubic unit cells ($4096$ atoms), employing geometry optimization to allow for the lattice to accommodate single or multiple S impurities. 
We include nine \textit{in}-\textit{situ}-optimised non-orthogonal local orbitals (with radius $10$ Bohr radii) for each site (in atomic Si, atomic orbitals are occupied up to level $3p$; for better convergence we additionally consider the five $3d$ orbitals) and use the PBE exchange-correlation functional \cite{Perdew1996} and a psinc grid with an 800 eV plane-wave cutoff \cite{Skylaris2002}. This gives an accuracy equivalent to plane-wave DFT for Si and other materials \cite{Skylaris2007}.
When embedded in silicon, sulfur, like the other chalcogens, acts as a deep donor.
Such defects have highly localized potentials that are well-described in a local orbital basis \cite{Yu2010}. 
The impurity distribution is generated by randomly substituting the impurity atoms onto  lattice sites. This follows the experimental techniques used to achieve high S concentrations, combining ion implantation with nanosecond pulsed-laser melting and rapid resolidification \cite{Winkler2011}. The impurities are effectively trapped in the substitutional sites \cite{Winkler2009}. 

The resulting Hamiltonians and overlap matrices, represented in terms of the nonorthogonal local orbital basis $\phi_a$,
are used to construct three catalogs of local Hamiltonian blocks (cf.\ Fig.\ \ref{fig:scheme}).  
The first catalog describes the Si host material, i.e., a set of onsite energies and hopping terms, starting at a central Si atom and extending to $10$ shells of Si neighbors. The second corresponds to the energies and hopping terms when the central atom is S, and the third catalog to pairs of neighboring S atoms. Here, we define a ``neighbor" as being at most $4$ shells apart. If two S atoms are $5$ or more shells apart, each S atom is unaffected by the presence of the other \cite{SM}.
For each system size $L$, concentration of impurities $n$ and disorder realization, we build the effective \emph{tight-binding} Hamiltonians $H$ and overlap matrices $O$ from these catalogs (cf.\ Fig.\ \ref{fig:scheme})  and 
%use the {\sc Jadamilu} implementation \cite{Schenk2006,Bollhofer2007} to 
solve the large generalized eigenvalue problem \cite{Schenk2006,Bollhofer2007}
\begin{equation}
H \psi_j = \varepsilon_j O \psi_j, \quad j= 1, \ldots, 9 L^3
\label{eq:GEV}
\end{equation}
for eigenenergies $\varepsilon_j$ and normalized eigenvectors $\psi_j= \sum_a M^a_j \phi_a$, written in a ``site'' basis by summing over the nine orbital coefficients of each site $k$, i.e.\ $|\psi_j(k)|^2 = \sum_{a \in k, b} M^{a}_j O_{ab} M^{b}_j$.
In Fig.\ \ref{fig:scheme}, we show examples of localized and extended states. 
For the $L^3=4096$ prototype, we have checked that our $\varepsilon_j$'s agree within $0.1\%$ -- $0.01\%$ with the DFT energy levels.
Due to the presence of $O$, and two orders of magnitude more hopping elements in $H$ compared to the Anderson model, we find that $10648$ atoms represent a practical upper limit (with tight-binding matrices of size $95832 \times 95832$). 
We average up to $1000$ different disorder realizations for each $L$ and $n$ (cf.\ Table \ref{tab:numerics}). 

%%%%%%%%%%%%%%%%%%%%%%%%%%%%%%%%%%%%%%%%%%%%%%%%%%%%%%%%%%%%
\begin{figure}
	\includegraphics[width=\columnwidth]{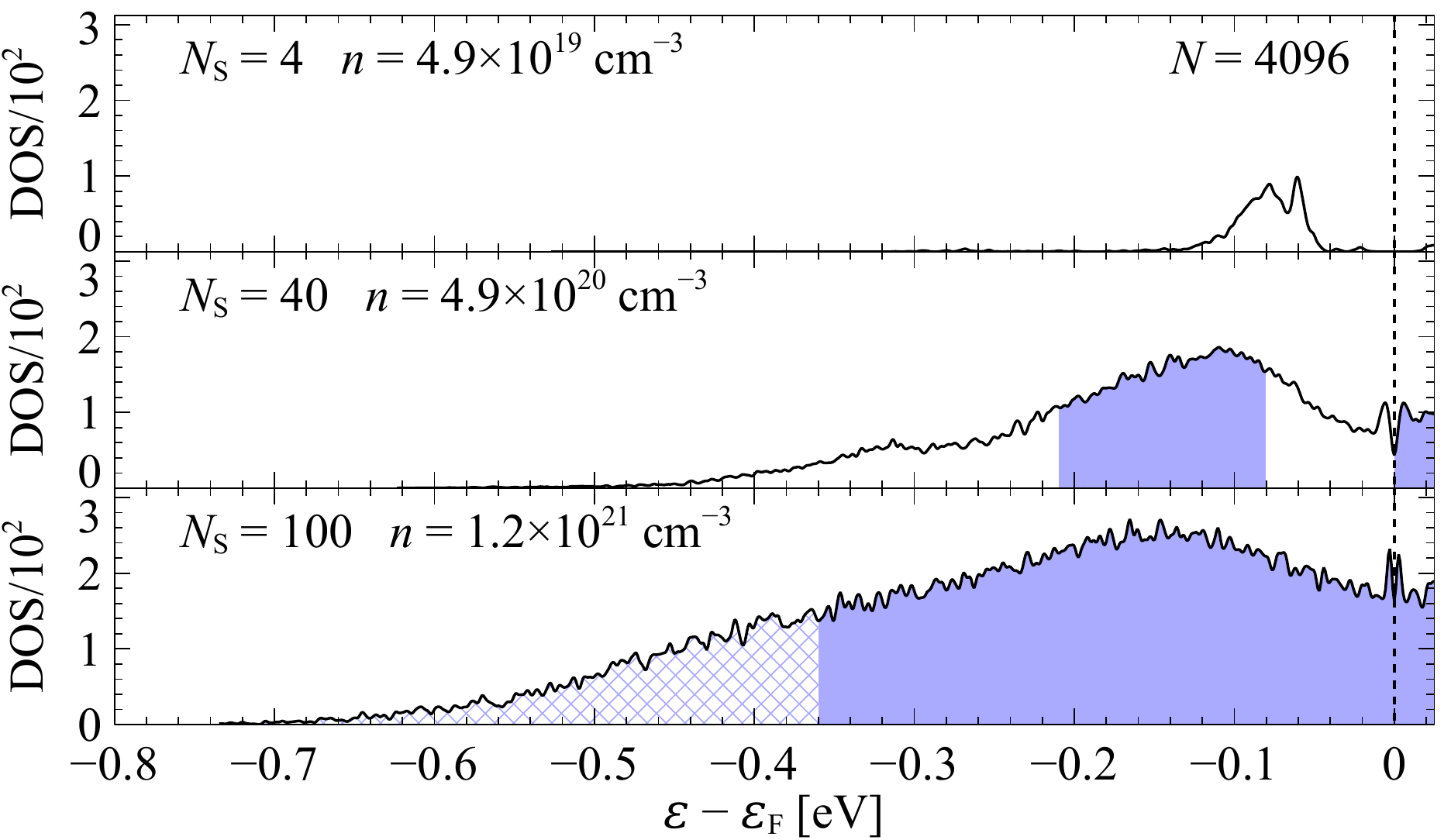}
	\caption{DOS of the IB for $ 4096 $ atoms at three different concentrations. 
		The shading indicates energies where states are on average delocalized (in the $L \rightarrow \infty$ limit), according to Fig.\ \ref{fig:PHDIAG}.  The delocalized CB states, separated by a vertical dashed line at $ \varepsilon_\text{F} $, are also shaded. Crossed shading indicates states that might be delocalized, but are outside our concentration range.}\label{fig:DOS}
\end{figure}
%%%%%%%%%%%%%%%%%%%%%%%%%%%%%%%%%%%%%%%%%%%%%%%%%%%%%%%%%%%%
%%%%%%%%%%%%%%%%%%%%%%%%%%%%%%%%%%%%%%%%%%%%%%%%%%%%%%%%%%%%
\begin{table}
	\caption{Summary of the range of impurities $N_\text{S} $, the concentration $ n $, the average, minimum, maximum and total number of disorder realizations for each $ L $, indicated by $ \langle \mathcal{N}\rangle $, $ \mathcal{N}_\text{min} $, $ \mathcal{N}_\text{max} $ and $ \sum \mathcal{N} $, respectively. The final column indicates the total number of $ \psi_j$'s per system size, and the last row the total for all $ N=L^3 $ and $ n $.}
	\begin{ruledtabular}
		\begin{tabular}{ccccccc}
			$N$	& $N_\text{S}$		& $ n/\SI{E20}{cm^{-3}} $	& $ \langle \mathcal{N} \rangle $ &	$(\mathcal{N}_\text{min},\mathcal{N}_\text{max})$ & $ \sum \mathcal{N} $ & $ \psi_j $'s	\\ \hline
			$16^3$	& 4--200  & 0.49--24  & 802 &(200,1000) & $ \num{68153} $ & $ \num{2943811} $\\ 
			$18^3$	& 5--322 & 0.43--28 &   758 & (106,1000) & $ \num{64430} $ & $ \num{3951351} $\\
			$20^3$	& 5--365 & 0.31--23 &  732 & (162,1000) & $ \num{71051} $ & $ \num{5640229} $\\ 
			$22^3$	& 10--410 & 0.47--19 &  541 & (293,733) & $ \num{34067} $ & $ \num{5521425} $ \\
			\multicolumn{5}{l}{Total no.\ of realizations and wave functions:} & $ \num{237311} $ & $ \num{18056816} $ \\
		\end{tabular}
	\end{ruledtabular}
	\label{tab:numerics}
\end{table}
%%%%%%%%%%%%%%%%%%%%%%%%%%%%%%%%%%%%%%%%%%%%%%%%%%%%%%%%%%%%
%
Characterizing the IB and its density of states (DOS) is interesting for its spin and charge transport properties \cite{Luque1997a,Wellens2016}. 
We compute the DOS of the IB from the $\varepsilon_j$'s while changing the number of impurities $N_\text{S}$. 
We define $\varepsilon_\text{F}$ as the midpoint between the highest occupied IB state at energy $\varepsilon_\text{IB}$ and the lowest unoccupied CB state at $\varepsilon_\text{CB}$. To obtain the \emph{average} DOS for given $N_\text{S}$ and $L$, we shift the spectrum of each realization such that $\varepsilon_\text{F}=0$.  The DOS shown in Fig.\ \ref{fig:DOS} is calculated by summing over Gaussian distributions of standard deviation $ \sigma = \SI{0.05}{mHa} = \SI{1.36}{meV} $ centered on $\varepsilon_j-\varepsilon_\text{F}$.
We find that the IB has a peak at {$\varepsilon - \varepsilon_\text{F}\sim \SI{-0.1}{eV}$} and a tail extending towards the VB with increasing $n$. This agrees with known features of the IB in doped semiconductors \cite{Cyrot-Lackmann1974}.
We emphasize that Si:S is particularly interesting for intermediate-band photovoltaic devices, where the efficiency increases when deep IB states can capture low-energy photons \cite{Luque1997a}. In order to avoid electron-photon recombination, the IB states should be delocalized such that they can contribute to the photocurrent. The determination of $n_\text{c}$ and the pronounced tail of the IB as presented in Fig.\ \ref{fig:DOS} therefore provide essential information for future device applications.

In the last decade, {multifractal analysis} \cite{Janssen1994,Milde1997,Evers2008} has become the method of choice to reliably and accurately extract the localization properties from wave functions \cite{Faez2009,Richardella2010,Rodriguez2011}. In its essence, it describes the scaling of various moments of the spatial distribution of  $|\psi_j|^2$, which is encoded in the \emph{singularity strengths} $\alpha_q$. 
We coarse grain $|\psi|^2$ \footnote{We drop the eigenstate index $j$ from here on.} by fixing a box size $l < L$ and partitioning the domain in $(L/l)^3 = \lambda^{-3}$ boxes. The amplitudes of the coarse-grained wave function $\mu$ are given by $\mu_s=\sum_{k \in B_s} |\psi(k)|^2$, i.e.\ by summing all $\vert \psi(k) \vert^2$ pertaining to the same box $B_s$. After rescaling the amplitudes as $ \log  \mu_s/\log \lambda $, we compute their arithmetic mean $ \alpha_0 = \braket{\log \mu_s}_s/\log \lambda $ and weighted mean $ \alpha_1 =  \sum_s \mu_s \log \mu_s /\log \lambda $ (proportional to the von Neumann entropy \cite{Cha10}). Finally, for each $ n $ and $ L $ we take the ensemble average $ \alpha_q(n,L) $, where $ q = 0 $ or $ 1 $.

At criticality, the universality class of the transition determines the scaling of $\alpha_q$ with $n$ and $L$. We capture this behavior using the well-established framework of finite-size scaling.
Following \cite{Rodriguez2011}, we assume that the data for each $ L $ meet at the critical point $ w = 0 $ with a value $ \alpha_q^{\text{crit}} $, and scale polynomially with $ \rho L^{1/\nu} $, where $\rho(w) = w + \sum_{m=2}^{m_\rho} b_m w^m$ includes higher-order dependencies on the dimensionless concentration $w = (n-n_\text{c})/n_\text{c}$. We hence fit the data against the function \footnote{{Note that we employ Wegner's relation \cite{Wegner1976} in assuming that in 3D the localization exponent equals the conductivity exponent. Although strictly speaking a hypothesis for interacting systems, it is still widely used in this context.}}
\begin{equation}
\alpha_q(n,L) = \alpha_q^\text{crit} + \sum_{i=1}^{m_L} a_i \rho^{i} L^{i/\nu} \, ,
\label{eq:fixed-scaling}
\end{equation}
with $ n_\text{c} $, $ \nu $, $ \alpha_q^\text{crit} $, the $ a_i $'s, and the $ b_i $'s as fitted parameters, and $ m_L $ and $ m_\rho $ as expansion orders \cite{SM}. 
We illustrate the localization and scaling properties of the wave functions using the moments $\alpha_0$ and $\alpha_1$.
Figure \ref{fig:PHDIAG} shows the results of the fits as $\varepsilon$ is varied, obtained from \eqref{eq:fixed-scaling} (see tables 1 %\ref{tab:FITPARAM-half} 
and 2 %\ref{tab:FITPARAM-quarter}
\cite{SM}). %
Crucially, we only accept estimates of $n_c$ and $\nu$ after consistently and rigorously checking their robustness against perturbations in $n$ and stability when increasing $m_L$ and $m_{\rho}$ \cite{Slevin1999,Rodriguez2011,SM}.

Following this recipe, we identify the Anderson MIT and reconstruct the energy dependence of the \emph{mobility edge} $n_\text{c}(\varepsilon)$ in Si:S. It exhibits (i) a maximum close to $\varepsilon_\text{F}$ and a decrease until $\varepsilon - \varepsilon_\text{F} \approx \SI{-0.09}{eV}$. (ii) For lower energies, $n_\text{c}$ increases again and the mobility edge moves towards the tail of the IB (cf.\ Fig.\ \ref{fig:DOS}).
These findings suggest a natural split into two different regimes, as also seen in the energy dependence of $\nu$. 
Values of $\nu$ in regime (i) increase continuously from $\nu \approx 0.5$ at $\varepsilon_\text{F}$ to about $\nu \sim 1$. In regime (ii), we find a larger spread of values $1 \lesssim \nu \lesssim 1.5$. This spread is consistent with the statistical uncertainty of each estimated $\nu$ in Fig.\ \ref{fig:PHDIAG}, which is dominated by the range of $L$ and the ensemble size $\mathcal{N}$ (cf.\ Tab.\ \ref{tab:numerics}). 
However, the trend in $\nu$ observed in regime (i) requires a different explanation.

\begin{figure}
	\includegraphics[width=\columnwidth]{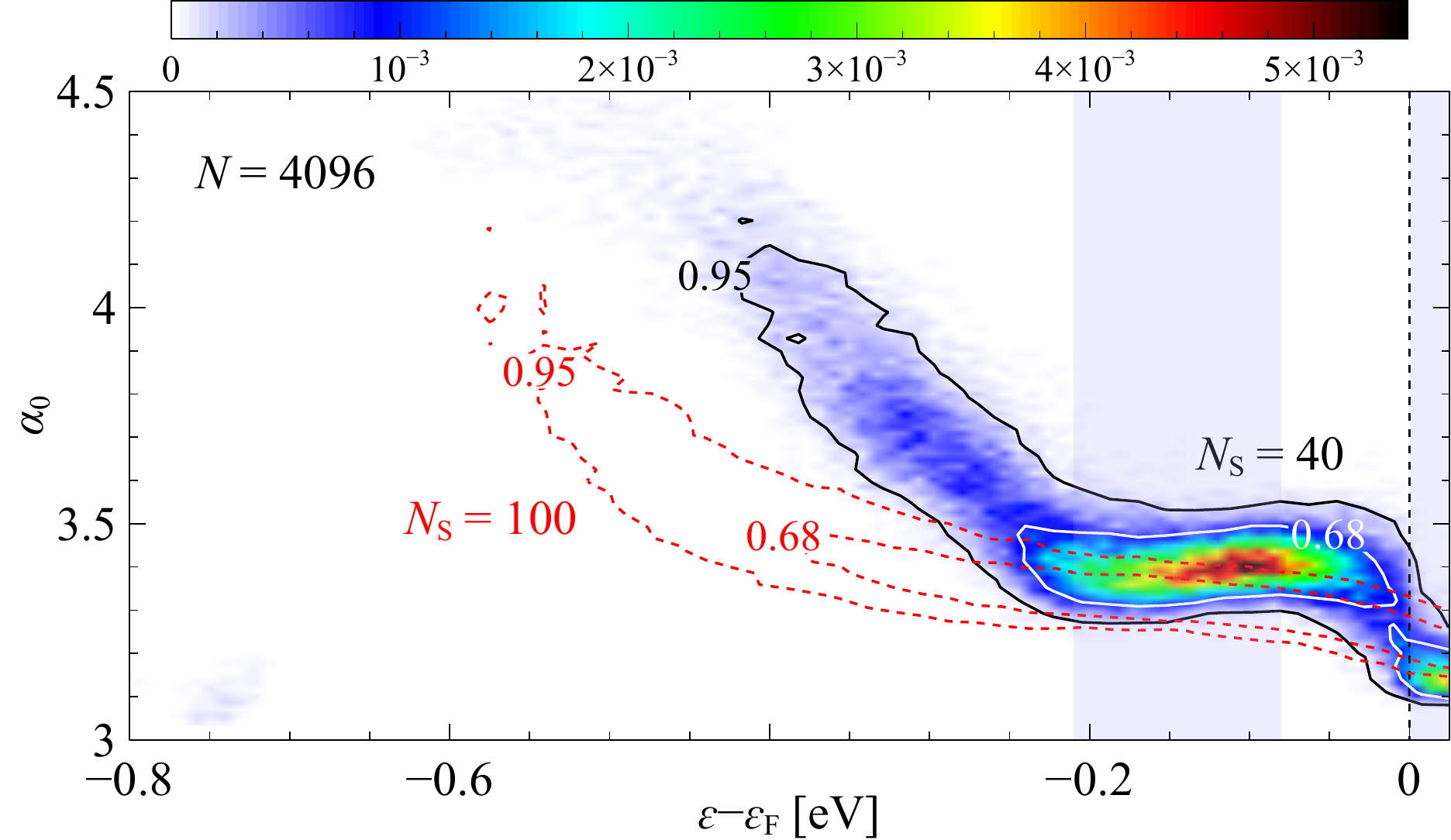} 
%     (b)\includegraphics[width=\columnwidth]{flat_bars.jpg} 
	\caption{Distribution of the moments $ \alpha_0 $ as a function of $\varepsilon$ shifted with $\varepsilon_\text{F}$ (vertical dashed line). For $N_\text{S} = 40$ we show the density plot of the distribution (from blue for low to red for high density, see color scale) and the contour lines enclosing 68\% (white) and 95\% (black) of the $ \alpha_0 $'s. For $N_\text{S} = 100$ we indicate the same contours (red, dashed). As in Fig. \ref{fig:DOS},  the shading denotes the delocalized region (in the $L \rightarrow \infty$ limit) according to Fig.\ \ref{fig:PHDIAG}.} 
	\label{fig:alpha_energy}
\end{figure}

\begin{figure}
    \includegraphics[width=\columnwidth]{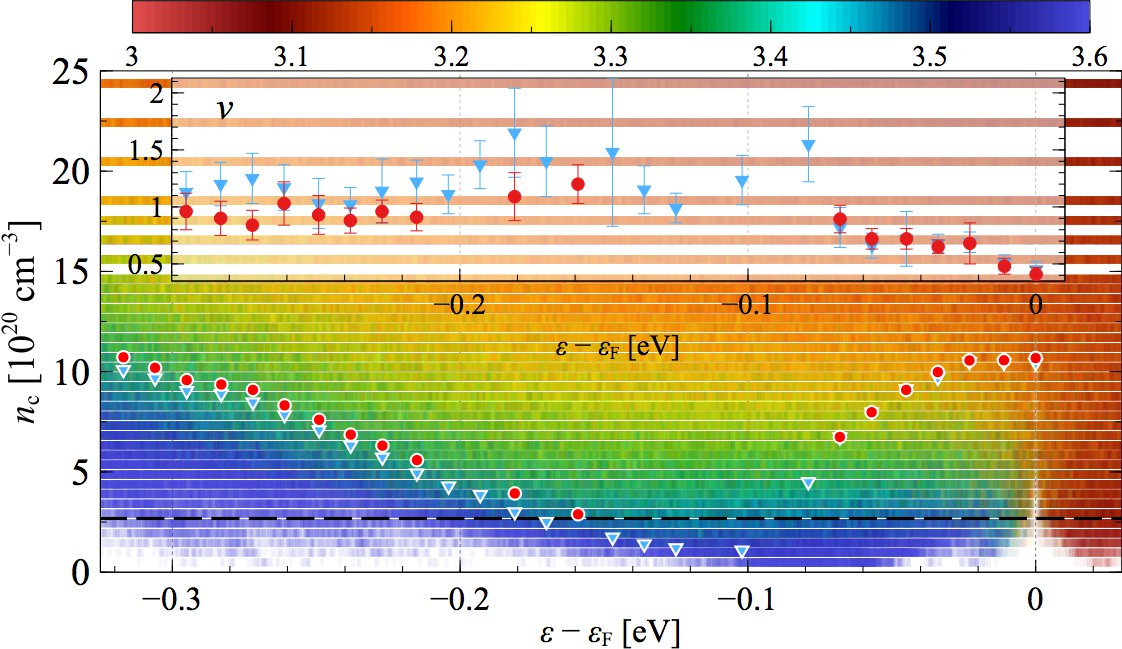} 
	\caption{{Moments $\alpha_0 (\varepsilon, n)$ for $N_\text{S} = 40$ indicated by colored bars as given in the color scale.
    In addition, we show the estimated $n_c$ from Fig.\ \ref{fig:PHDIAG} for $\lambda=1/4$ with $q=0$ ($\circ$) and $1$ ($\triangledown$). The horizontal gray line at $n\lesssim 3$ indicates the lack of enough low-doping S concentrations to allow scaling fits for $q=0$. 
Regions of $\alpha_0 \sim 3$, indicating much less-localised states, begin to extend below the Fermi level at concentrations above where the gap closes.
Inset: For easier comparison, we plot the estimated $\nu$ values for $\lambda=1/4$ with $q=0$ ($\circ$) and $1$ ($\triangledown$) as in Fig.\ \ref{fig:PHDIAG}. }
}
	\label{fig:alpha_energy_concentration}
\end{figure}

In Fig.\ \ref{fig:alpha_energy}, we present the distribution of states resolved in both energy $ \varepsilon$ and $\alpha_0$.
Perfectly extended states correspond to $\alpha_0 = 3$, while increasing localization results in $\alpha_0 \rightarrow\infty$.
%%%%%%%%%%%%%%%%%%%%%%%%%%%%%%%%%%%%%%%%%%%%%%%%%%%%%%%%%%%%
%%%%%%%%%%%%%%%%%%%%%%%%%%%%%%%%%%%%%%%%%%%%%%%%%%%%%%%%%%%%
The data for $N_\text{S}=40$ ($n=\SI{4.9E20}{cm^{-3}}$) show metallic states of the CB with $\alpha_0 \approx 3$ at $\varepsilon \approx \varepsilon_\text{F}$. The IB is characterized by (i) a majority region of states with $\alpha_0 \approx 3.4$ for $(\varepsilon-\varepsilon_\text{F}) \in [\SI{-0.25}{eV},\SI{-0.05}{eV}]$ and (ii) a tail region of more localized $\alpha_0$ values ($\geq 3.4$) for $(\varepsilon-\varepsilon_\text{F}) \lesssim \SI{-0.25}{eV}$. However, for $N_\text{S} = 100$ ($n=\SI{1.2E21}{cm^{-3}}$) IB and CB have lost their identities.
In fact, the two bands overlap in $\varepsilon$ as shown in Fig.\ \ref{fig:DOS}, and also change their localization properties --- the bands have hybridized, with $\alpha_0$ decreasing towards $3$ (the metallic limit) close to $\varepsilon_\text{F}$.
This observation is intriguing when tensioned against the simultaneous decrease in the value of $\nu$ at $\varepsilon \sim \varepsilon_\text{F}$ (cf.\ Fig.\ \ref{fig:PHDIAG}). {Apparently the localization of the IB states is substantially modified by the presence of the states from the CB.}
{
In Fig.\ \ref{fig:alpha_energy_concentration}, we show the 
$\alpha_0$ data for $N=4096$ as a function of $\varepsilon$ and $n$. For small impurity concentrations, the IB consists of localized states with some of the largest values of $\alpha_0\sim 3.6$, while the CB contains delocalized states with $\alpha_0\gtrsim 3$.
Upon increasing $n$, the IB develops and its states become more delocalized. Initially, this trend is most pronounced where the DOS of the IB is large (see Fig.\ \ref{fig:DOS}), i.e.\ around $\varepsilon - \varepsilon_\text{F}\sim \SI{-0.12}{eV}$. Simultaneously, states at the top of the IB exhibit $\alpha_0$ values close to those denoting extended states in the CB, even before the band gap has fully closed.
When reliable scaling is possible, we eventually see how the two mobility edges emerge. At the lower mobility edge, we find values of $\nu \sim 1-1.5$. At the upper mobility edge, we observe 
lower estimates for $\nu$ coinciding with lower $\alpha_0$ values at the transition due to the strong hybridization of IB and CB. 
}
%
%%%%%%%%%%%%%%%%%%%%%%%%%%%%%%%%%%%%%%%%%%%%%%%%%%%%%%%%%%%%
%%%%%%%%%%%%%%%%%%%%%%%%%%%%%%%%%%%%%%%%%%%%%%%%%%%%%%%%%%%%
Let us discuss how this observed hybridization and the resulting enhanced metallic behavior
can affect the value of $\nu$. The leading scaling behavior from \eqref{eq:fixed-scaling} is $\alpha_0^\text{crit} - \alpha_0 \sim w L^{1/\nu}$ for $ w > 0 $. A decrease in the effective $\alpha_0$ yields an increase in  $\alpha_0^\text{crit} - \alpha_0$, which is consistent with a \emph{reduced} exponent $\nu$ as observed in Fig.\ \ref{fig:PHDIAG} for $(\varepsilon-\varepsilon_\text{F}) \gtrsim \SI{-0.1}{eV}$. An argument similar to the famous ``gang of four'' result \cite{Abrahams1979} can be made directly for the transport experiments, where an \emph{increase} in $\sigma\sim w^{\nu}$ for $0 \leq w \ll 1$, i.e.\  close to the critical point, is also consistent with a reduced $\nu$. 

%%%%%%%%%%%%%%%%%%%%%%%%%%%%%%%%%%%%%%%%%%%%%%%%%%%%%%%%%%%%%
%\section*{Outlook}\label{sec-outlook}
%%%%%%%%%%%%%%%%%%%%%%%%%%%%%%%%%%%%%%%%%%%%%%%%%%%%%%%%%%%%

Let us reiterate our main point: our simulations of the Anderson MIT in an uncompensated doped semiconductor find an effective $\nu\approx 0.5$ near $\varepsilon_\text{F}$, where the IB and the CB hybridize. 
Larger values of $\nu$, around $1$ to $1.5$, can be observed when the level of compensation is increased. These results provide a possible explanation for the observation of Itoh et al.\ \cite{ItoWOH04} that in experiments a change from $0.5$ to $\nu \sim 1$ can be induced by compensation. Taken together, compensation and band hybridization provide two important pieces to complete the ``exponent puzzle''{: modelling the Anderson transition in doped semiconductors needs to include the CB (VB for hole-doped materials) together with the IB provided by the Anderson model --- the experiments obviously include both and hence find $\nu$ values which, depending on their state of compensation, can be different from predictions based solely on the Anderson model of the IB.}

{How exactly the hybridization changes the effective value of $\nu$, as well as whether the value of $\nu$ deep in the IB is different from the non-interacting predictions, remain challenges for future high-precision studies.}
This includes understanding how the hybridization of two bands, each with its own characteristic length scale, fits the hypotheses of Chayes' theorem \cite{Chayes1986}.

Still, the approach we present here exploits and transfers the accuracy and versatility of modern \emph{ab initio} simulations to the study of Anderson localization in doped semiconductors---at a fraction of the computational cost.
Beyond bulk semiconductors, other disordered systems \cite{Chakraborty2017}, 2D \cite{Geim2007,Bhimanapati2015,Das2015a} and layered materials \cite{Wilson2017} are also well within reach, as is the investigation of the influence of many-body physics by, e.g., studying the interaction-enabled MIT in 2D Si:P \cite{Abrahams2001,Punnoose2005}.
We find that the critical concentration agrees quantitatively with a previous experiment in Si:S by Winkler et al.\ \cite{Winkler2011}. 
Our approach is hence capable of modeling fundamental physical phenomena while also making material-specific predictions. 

%%%%%%%%%%%%%%%%%%%%%%%%%%%%%%%%%%%%%%%%%%%%%%%%%%%%%%%%%%%%
%\subsection*{Acknowledgments}
	We are grateful to Amnon Aharony, Siddharta Lal, David Quigley and Alberto Rodriguez for discussions. We thank EPSRC for support via the ARCHER RAP project e420 and the MidPlus Regional HPC Centre (EP/K000128/1). UK research data statement: data is available at \footnote{\href{http://wrap.warwick.ac.uk/id/eprint/92910}{http://wrap.warwick.ac.uk/id/eprint/92910}}.

% %%%%%%%%%%%%%%%%%%%%%%%%%%%%%%%%%%%%%%%%%%%%%%%%%%%%%%%%%%%%

%%%%%%%%%%%%%%%%%%%%%%%%%%%%%%%%%%%%%%%%%%%%%%%%%%%%%%%%%%%%
\bibliographystyle{prsty}

%%%%%%%%%%%%%%%%%%%%%%%%%%%%%%%%%%%%%%%%%%%%%%%%%%%%%%%%%%%%%

\clearpage

\part{Supplemental Materials}

	In these Supplemental Materials we provide details on the DFT simulations of the prototypes used to build the catalogs of matrix elements. We compare the results obtained from ONETEP and the effective model of an exemplary system of 4096 atoms with 29 impurities. We then show the scale-invariant distribution of the multifractal exponents $ \alpha_0 $ of the wave function, a hallmark of the metal-insulator transition. Finally, we discuss the scaling behaviour of $ \alpha_0 $ with concentration and system size to obtain estimates of the critical parameters of the transition.

%\twocolumngrid

%%%%%%%%%%%%%%%%%%%%%%%%%%%%%%%%%%%%%%%%%%%%%%%%%%%%%%%%%%%%
\section{DFT simulations}
%%%%%%%%%%%%%%%%%%%%%%%%%%%%%%%%%%%%%%%%%%%%%%%%%%%%%%%%%%%%

Simulations employ the ONETEP linear-scaling DFT package \cite{Skylaris2005}, which describes the single-electron density matrix via local orbitals (``nonorthogonal generalized Wannier functions'', denoted NGWFs) and a kernel matrix, both optimized \textit{in situ} \cite{Skylaris2005}.
We use the PBE exchange-correlation functional {\cite{Perdew1996}} and include all nine orbitals (\textit{s}, \textit{p} and \textit{d}) for each site, with a cutoff radius of $10$ Bohr radii for each NGWF, which are described using psinc functions on a grid set by a 800 eV plane-wave cutoff. This combination of methods has been shown to deliver accuracy equivalent to plane-wave DFT for Si and other materials \cite{Skylaris2007}.
We note that our chosen system size of $4096$ atoms is large enough to avoid the interaction of defect states with their periodic images.
These calculations result in DFT energy levels $\varepsilon^\mathrm{DFT}_j$ and states $\psi^\text{DFT}_j$ as well as the determination of the energy $\varepsilon^\text{DFT}_\text{VB}$ of the top of the valence band (VB) and the energy $\varepsilon^\text{DFT}_\text{CB}$ of the bottom of the CB. The states $\psi^\text{DFT}_j$ are expanded in a basis of NGWFs $\phi_\alpha$ \cite{Skylaris2005}, with associated overlap matrix $O_{\alpha\beta}=\langle \phi_\alpha | \phi_\beta \rangle$.

\begin{figure}[t]
	\includegraphics[width=\columnwidth]{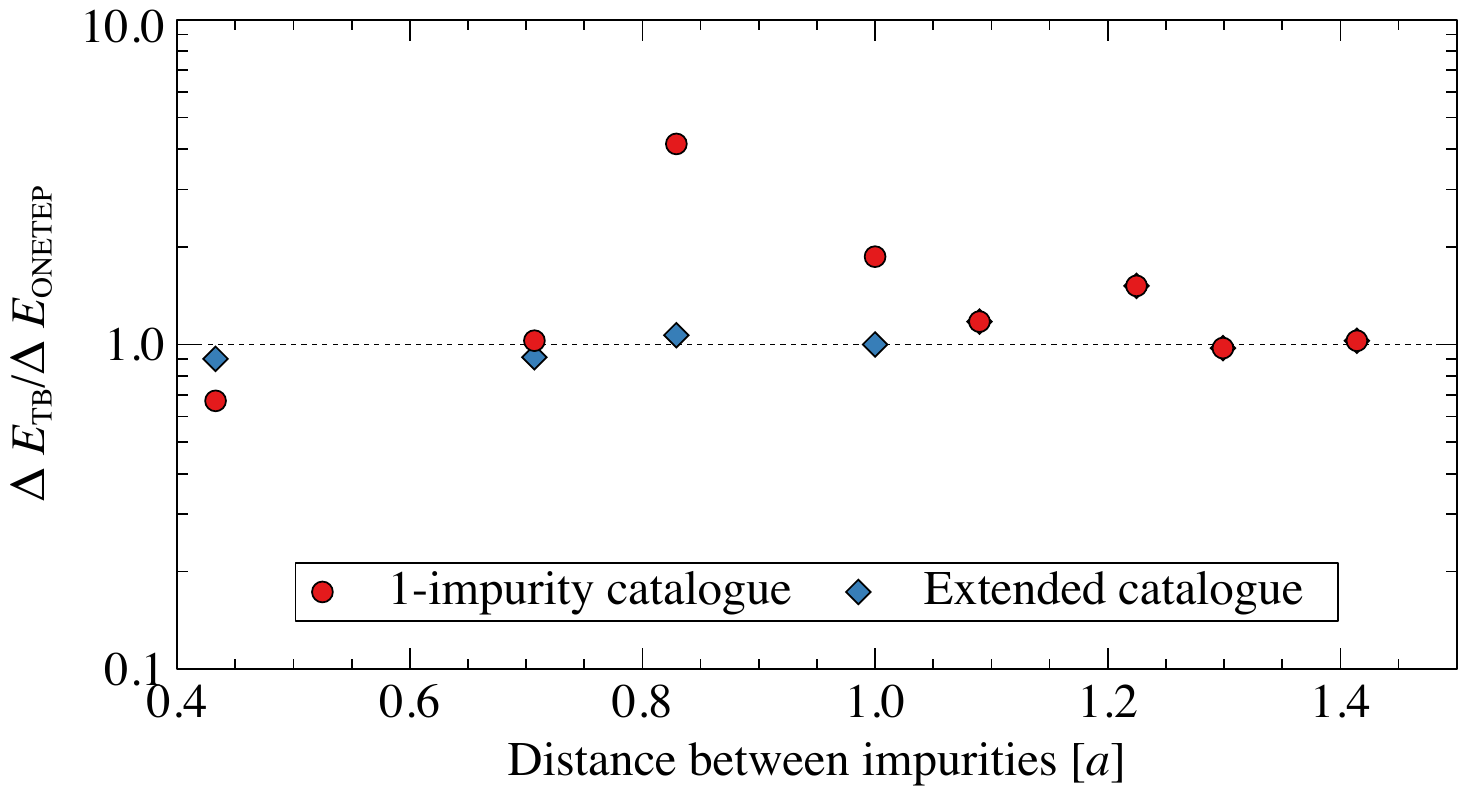}
	\caption{Ratio {$\Delta E_\text{TB}/\Delta E_\text{ONETEP}$ (see text for definition)
			of the energy difference} between the two impurity states appearing when a system of $ \num{4096} $ atoms is doped with a pair of defects at increasing distance. The ``1-impurity catalogue'' is built from systems with only one impurity, while the ``extended catalogue'' includes the description of pairs of defects.
		%For situations where only one state in the gap appears, discussed in the text, the gap is taken between said state and the lowest conduction band state.
		\label{fig:pair_catalogue}}
\end{figure}

When constructing the catalogs we account for the symmetries of the $p$ and $d$ orbitals in relation to the diamond-cubic structure of the Si lattice, in order to identify symmetrically \mbox{(in-)equivalent} configurations.
The third catalog includes the matrix element describing pairs of neighboring defects that are up to one unit cell apart {(the Si lattice constant is $a = \SI{10.1667}{Bohr}$ \cite{Skylaris2007})}. We have set this cut off by comparing the ONETEP runs of pairs at increasing distance to their effective tight-binding model using just the first and second catalogs. Because each defect induces a state in the band gap of Si, we compute the difference in energy between the defect states {in both ONETEP, $\Delta E_\text{ONETEP}$, and our effective tight-binding model, $\Delta E_\text{TB}$. In Fig.\ \ref{fig:pair_catalogue} we plot the ratio $\Delta E_\text{TB}/\Delta E_\text{ONETEP}$ for the two cases of defects treated as single impurities (``1-impurity catalogue'') and as a pair (``2-impurity catalogue'') --- for good agreement we expect $\Delta E_\text{TB}/\Delta E_\text{ONETEP} \approx 1$.}
When S defects are first nearest neighbors, they form bonding and anti-bonding states, with the former disappearing into the VB. In this case we compare the distance of the anti-bonding state to the lowest-energy CB state.
While the 1-impurity catalog manages to capture this feature well for first nearest neighbors, the corresponding 2-impurity catalogs gives a better description. Moreover, the 1-impurity catalog predicts a similar situation when S defects are second nearest neighbors. This contradicts the results from ONETEP and is correctly rectified in the extended catalog.

Since each S impurity donates two electrons to the system, and we assume double occupation of the energy levels in the impurity band (IB), the number of impurities $N_\text{S}$ coincides with the number of states in the band gap. States in the VB (CB) are extended and can be easily identified from their energy and participation ratio $P=1/(N \sum_{k} |\Psi_k|^4)$, assuming the convenient value $P > 0.3$ (suppressing the $j$ index). We check if two dopants are nearest-neighbors and subtract the number of missing bonding states from the number of IB states.

%%%%%%%%%%%%%%%%%%%%%%%%%%%%%%%%%%%%%%%%%%%%%%%%%%%%%%%%%%%%
%%%%%%%%%%%%%%%%%%%%%%%%%%%%%%%%%%%%%%%%%%%%%%%%%%%%%%%%%%%%

\begin{figure}[t]
	\includegraphics[width=0.95\columnwidth]{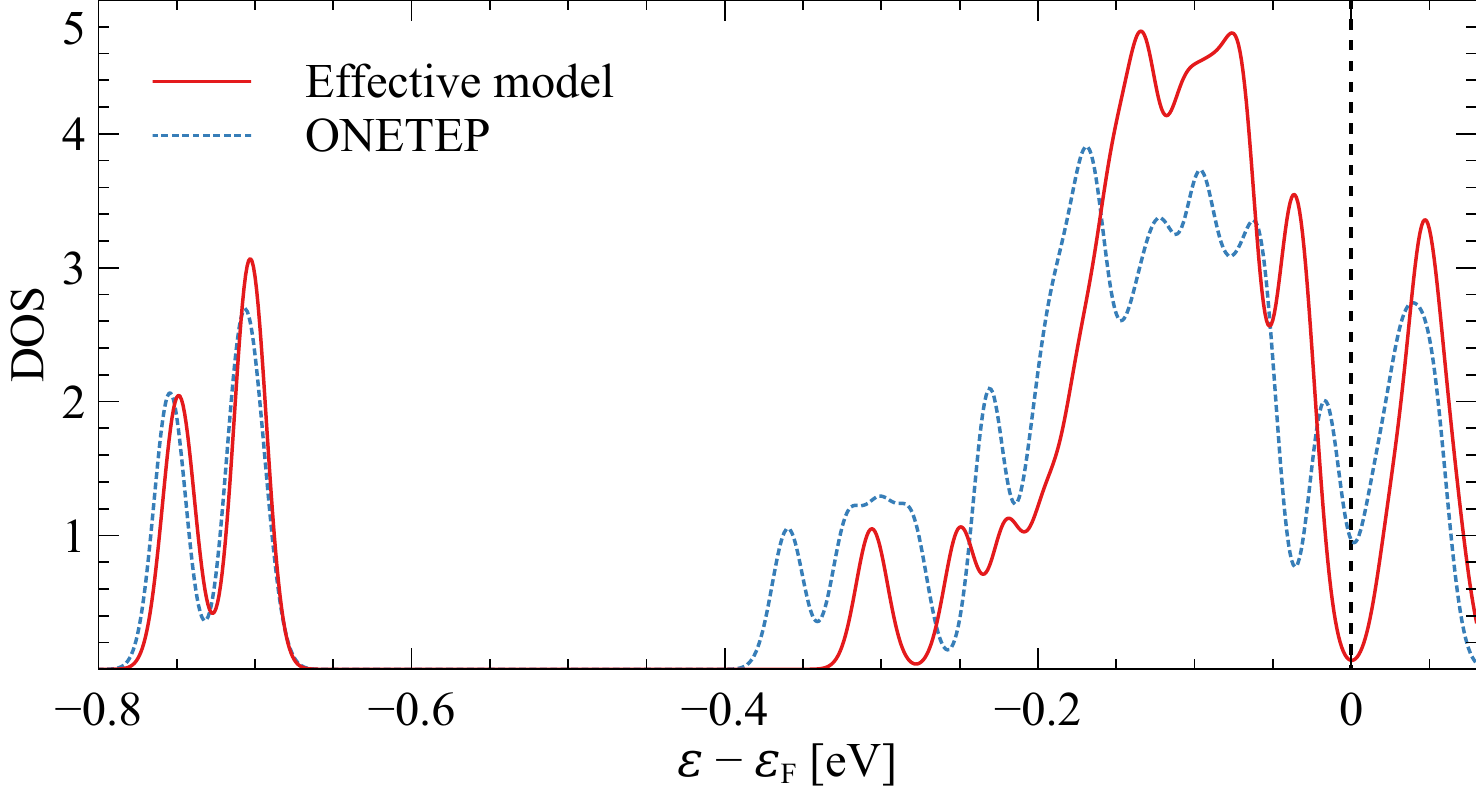}
	\caption{DOS of the exemplary system of Fig.\ 2 of the main text ($ 4096 $ atoms, $ 29 $ impurities). In blue dashed we show the spectrum obtained from the ONETEP simulation, in red that from its effective model. The Gaussian broadening used is $ \sigma = \SI{10}{meV} $. The spectra are aligned at the Fermi energy $ \varepsilon_\text{F} $, indicated by the dashed line. \label{fig:DOS_compare}}
\end{figure}

In Fig.\ \ref{fig:DOS_compare} we compare the spectrum of a $ 4096 $-atom system with 29 impurity (shown in Fig.\ 2 of the main text) obtained from a ONETEP calculation and from the corresponding effective tight-binding model.
At least for this particular disorder realization, the position of the VB and CB qualitatively coincide in the two spectra, and the IB extends towards the center of the band gap and show a higher density of states closer to the Fermi energy. The number of impurity states is also correctly counted.

%%%%%%%%%%%%%%%%%%%%%%%%%%%%%%%%%%%%%%%%%%%%%%%%%%%%%%%%%%%%
%%%%%%%%%%%%%%%%%%%%%%%%%%%%%%%%%%%%%%%%%%%%%%%%%%%%%%%%%%%%

%%%%%%%%%%%%%%%%%%%%%%%%%%%%%%%%%%%%%%%%%%%%%%%%%%%%%%%%%%%%
\begin{figure}[t]
	\vspace{2ex}
	\includegraphics[width=0.95\columnwidth]{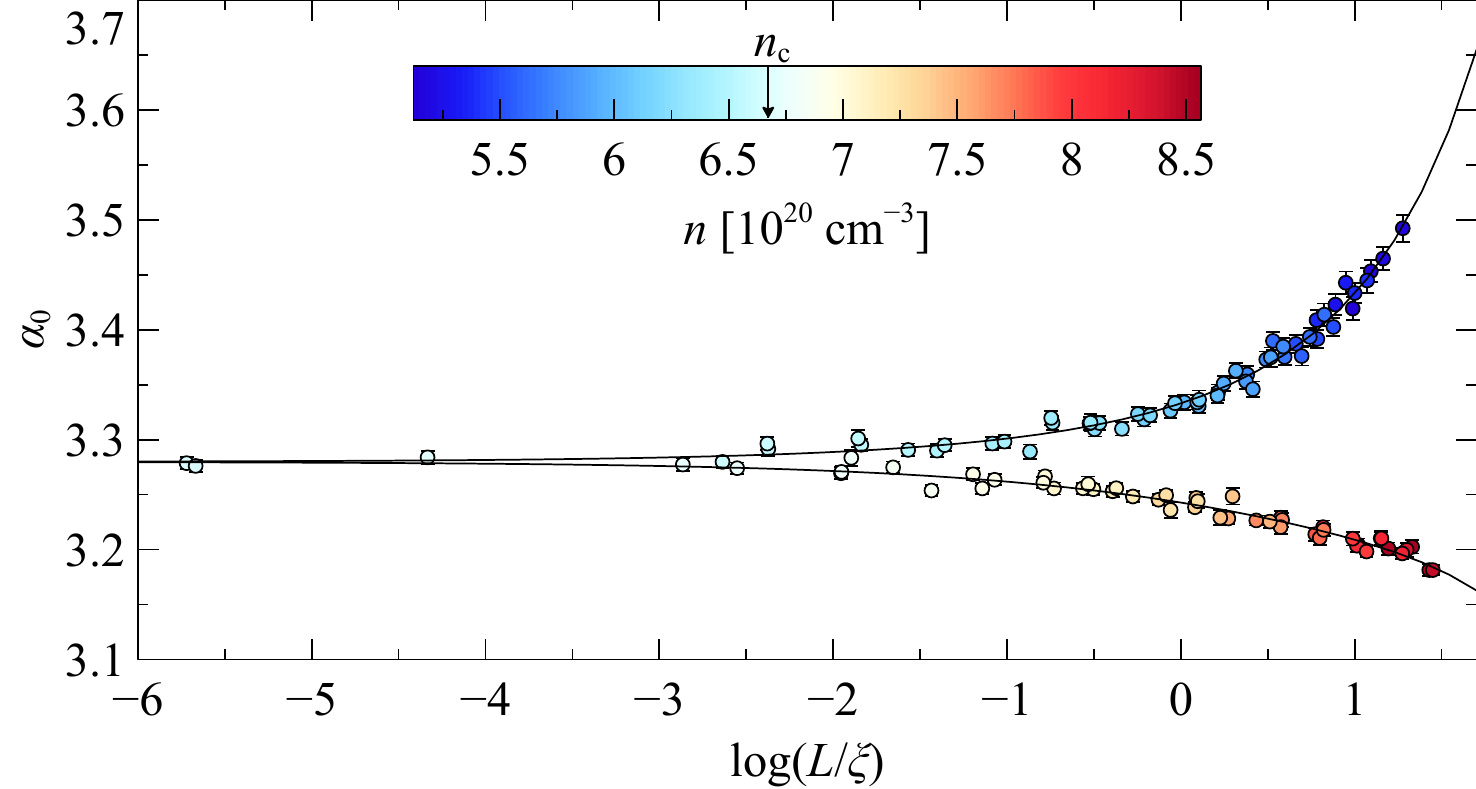}
	\caption{Scaling of $ \alpha_0 $ as a function of $ L/\xi $, defined in Eq.\ \eqref{eq:loc-length}, for energy $ -\SI{0.249}{eV} $. The points indicate the numeric values of $ \alpha_0 $ obtained from the ensemble average, with colors indicating the value of $ n $ for each data point. The critical concentration is indicated as $ n_\text{c} $ on the scale. The underlying fit is given by the function \eqref{eq:scaling-function}. \label{fig:scaling-example}}
\end{figure}
%%%%%%%%%%%%%%%%%%%%%%%%%%%%%%%%%%%%%%%%%%%%%%%%%%%%%%%%%%%%

%%%%%%%%%%%%%%%%%%%%%%%%%%%%%%%%%%%%%%%%%%%%%%%%%%%%%%%%%%%%
\section{Closure of the band gap}
%%%%%%%%%%%%%%%%%%%%%%%%%%%%%%%%%%%%%%%%%%%%%%%%%%%%%%%%%%%%

{To observe the MIT, the band gap $\varepsilon_\text{gap}= \varepsilon_\text{CB} - \varepsilon_\text{IB}$ must vanish.  
	Let $\Delta_\text{IB}$ denote the average energy level spacing in the IB. When $\varepsilon_\text{gap} \gg \Delta_\text{IB}$ the system has a gap and is non-metallic. Hence $\varepsilon_\text{gap} \lesssim \Delta_\text{IB}$ is necessary for metallic behavior to emerge upon increasing $N_\text{S}$. In Fig.\ \ref{fig:BG} we show that the gap closure indicator $\delta_\text{IB} = (\varepsilon_\text{gap} - \Delta_\text{IB})/\varepsilon_\text{gap} \approx 0$ at $n_0 \approx \SI{8E20}{cm^{-3}}$. For $N=4096$, the value corresponds to $n_0 \approx \SI{6.5E20}{cm^{-3}}$.
	%Only for higher concentrations we can expect a dense region of states and hence a MIT. Otherwise, we might still find that the states become delocalized, but, being occupied, they cannot contribute to transport.
	For $n \approx n_0$, the energy gap has closed and metallic behavior emerges when the wave functions become delocalized by further increasing $n$.
	% --- a necessary condition to observe the MIT.
	%When the wave functions become delocalized, metallic behavior can emerge. 
	%%%%%%%%%%%%%%%%%%%%%%%%%%%%%%%%%%%%%%%%%%%%%%%%%%%%%%%%%%%%
	\begin{figure}
		\includegraphics[width=\columnwidth]{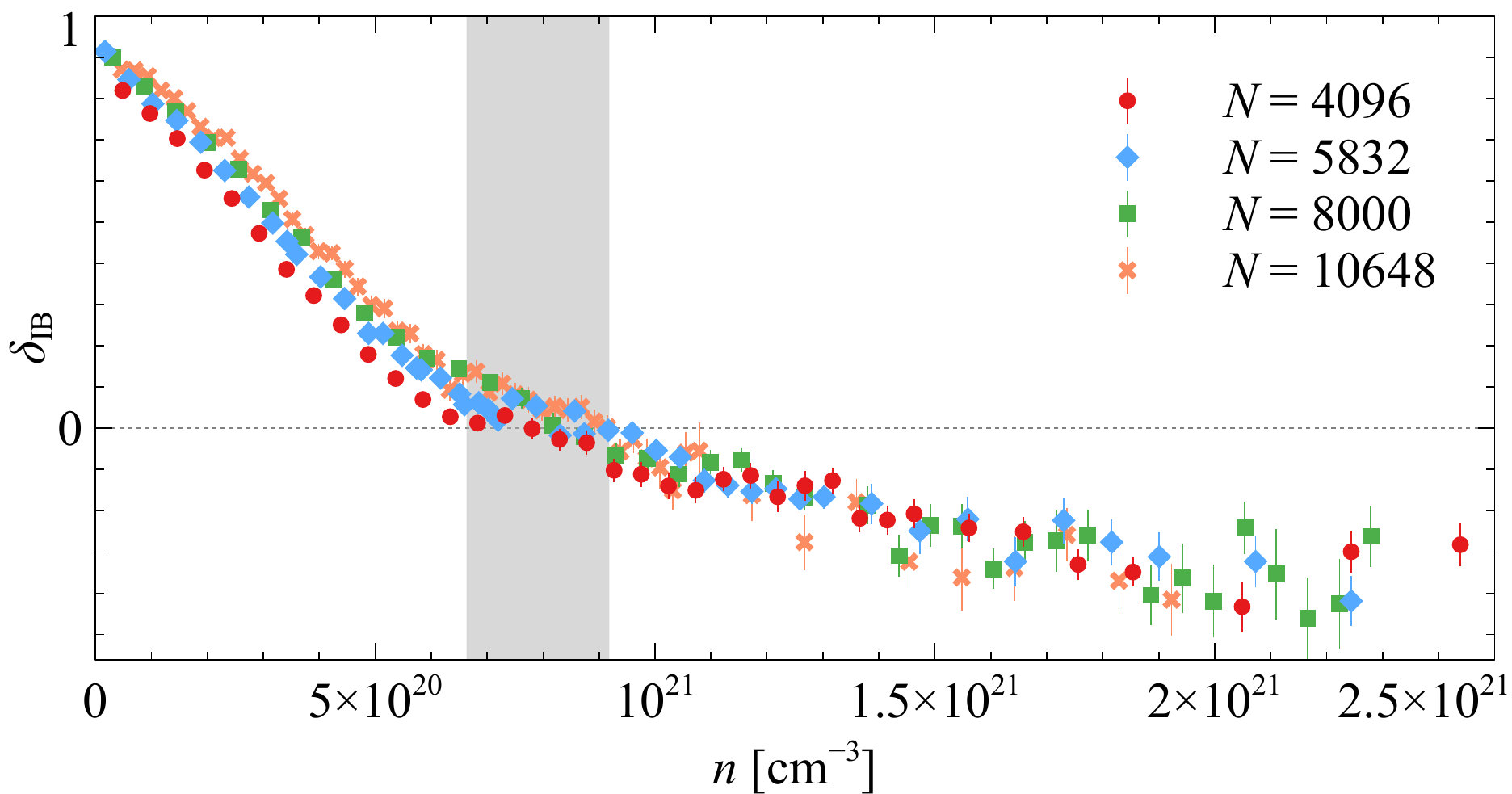} 
		\caption{
			Gap closure indicator $ \delta_\text{IB} $ 
			as a function of $n$. 
			Error bars indicate the standard error of the mean. The area shaded in gray highlights the concentrations at which IB and CB mix.}\label{fig:BG}
	\end{figure}
	%%%%%%%%%%%%%%%%%%%%%%%%%%%%%%%%%%%%%%%%%%%%%%%%%%%%%%%%%%%%
	Let us emphasize that our condition $\delta_\text{IB}\approx 0$ will likely overestimate the $n_0$ value due to the non-uniform distribution of the IB DOS. 
}

%%%%%%%%%%%%%%%%%%%%%%%%%%%%%%%%%%%%%%%%%%%%%%%%%%%%%%%%%%%%
\section{Invariance of the PDF at criticality}
%%%%%%%%%%%%%%%%%%%%%%%%%%%%%%%%%%%%%%%%%%%%%%%%%%%%%%%%%%%%

\begin{figure*}
	\includegraphics[width=\textwidth]{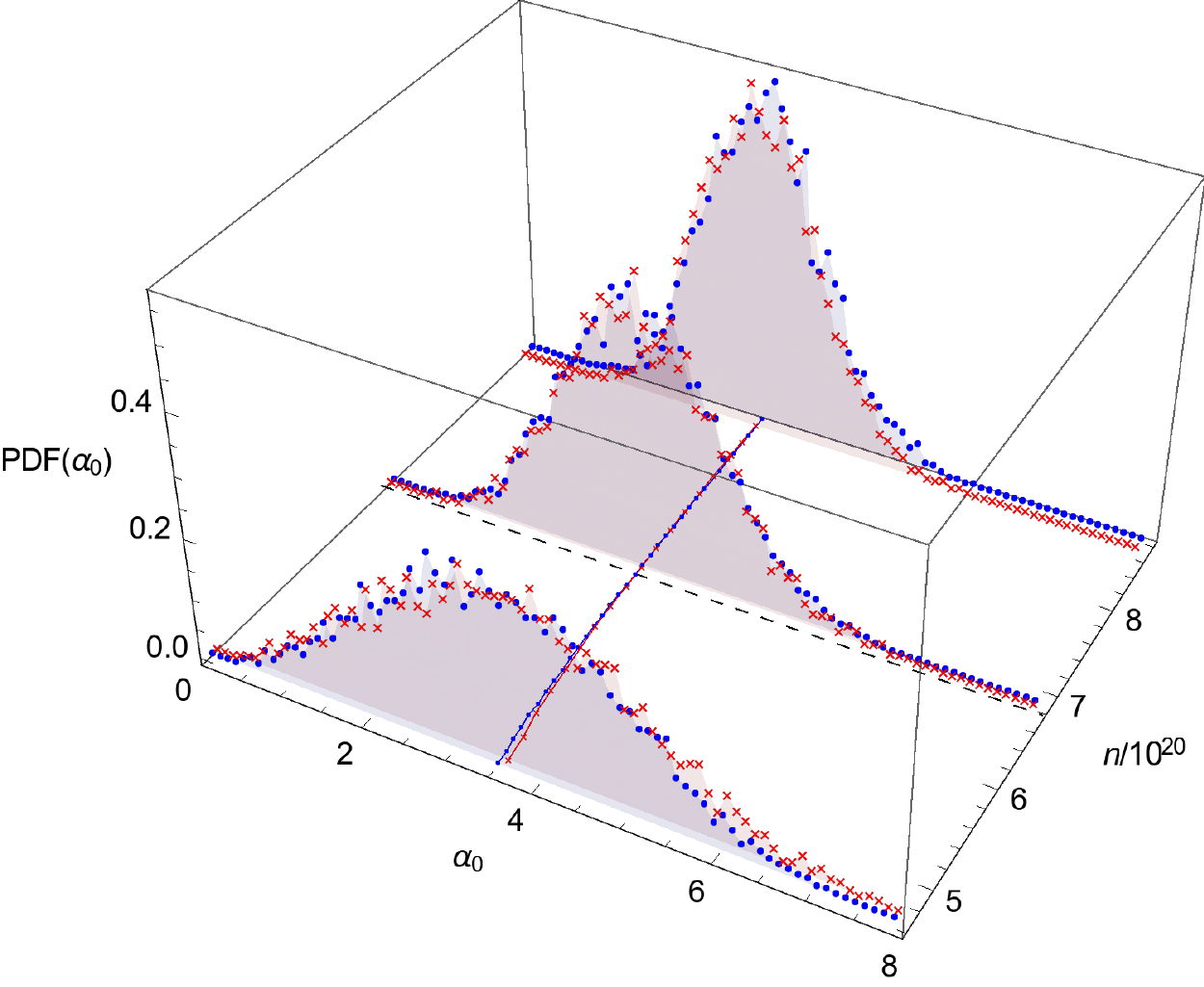}
	\caption{Distribution for the $ \alpha_0 $ exponent at coarse-graining $ \lambda = 1/2 $ and energy $ \varepsilon - \varepsilon_\text{F} = \SI{-0.249}{eV} $, as a function of the concentration $ n $ (in units of $ \SI{E20}{cm^{-3}} $), for two system sizes $ L^3=4096 $ (blue dots) and $ 10648 $ (red crosses). For clarity we show the histogram for three concentrations: before the transition ($ n = \SI{4.6E20}{cm^{-3}} $), near the critical point ($ \SI{6.8E20}{cm^{-3}} $), and after ($ \SI{8.8E20}{cm^{-3}} $). The critical point ($ n_\text{c} = \SI{6.7E20}{cm^{-3}} $) is indicated by a black dashed line and the value used is from table \ref{tab:FITPARAM-half}. On the bottom plane we show the position of the averages also for the intermediate concentrations, connected by lines to guide the eye. We use again blue dots with a solid line for $ L^3=4096 $ and red crosses with a dashed line for $ 10648 $.}
	\label{fig:pdf-invariance}
\end{figure*}

The probability distribution function $ P $ of the multifractal exponent $ \alpha_0 $ depends, at the critical point, only on the coarse-graining $ \lambda = l/L $, rather than separately on the system size $ L $ and the box size $ l $ \cite{Rodriguez2011}. This implies that, again at the critical point, the probability distribution of $ \alpha = \log \mu/\log \lambda $ ($ \mu $ is the coarse-grained wave function, refer to the main text) has the same shape for any $ L $, provided that the wave functions are coarse-grained with the matching $ l = \lambda L $ box size. The dependence of $ P(\alpha) $ on $ L $ gradually reappears away from the critical point, where in the localized (delocalized) regime larger systems become more localized (delocalized). This is shown in Fig.\ \ref{fig:pdf-invariance}, where we plot the $ P(\alpha) $ at $ \lambda=1/2 $ for three values of $ n $: the lowest in the localized regime, the intermediate close to the critical point and the highest in the delocalized regime. In addition, tables \ref{tab:FITPARAM-half} and \ref{tab:FITPARAM-quarter} report the results of the finite-size scaling analysis that are displayed in Fig. 1 in the main text.

%%%%%%%%%%%%%%%%%%%%%%%%%%%%%%%%%%%%%%%%%%%%%%%%%%%%%%%%%%%%
\section{Scaling of the multifractal exponents}
%%%%%%%%%%%%%%%%%%%%%%%%%%%%%%%%%%%%%%%%%%%%%%%%%%%%%%%%%%%%
The metal-insulator transition is observed in the power-law divergence of the conductance (susceptibility) in the metallic (insulating) regime, or, equivalently, in the correlation (localization) length
\begin{equation}\label{eq:loc-length}
\xi \propto \left| \frac{n-n_\text{c}}{n_\text{c}} \right|^{-\nu} \, . 
\end{equation}
Following the scaling theory of localization \cite{Kramer1993}, we assume that all relevant quantities near the transition are determined by the renormalized length $ L / \xi $. In Fig.\ \ref{fig:scaling-example} we plot $ \alpha_0 $ for different $ n,L $ as a function of $ \log(L/\xi) $ and show that these data collapse on a single curve, whose parameters are determined by fitting Eq.\ 2 from the main text. Specifically for Fig.\ \ref{fig:scaling-example}, the scaling function reads
\begin{align}\label{eq:scaling-function}
\alpha_q^{(\lambda)}(n,L) & = \alpha_q^\text{crit} + 
a_1 \frac{n-n_\text{c}}{n_\text{c}} L^{1/\nu} \nonumber\\
& +
a_2 \left( \frac{n-n_\text{c}}{n_\text{c}} L^{1/\nu} \right)^2 +
a_3 \left( \frac{n-n_\text{c}}{n_\text{c}} L^{1/\nu} \right)^3 \, .
\end{align}
Notice that this equation is a third-order polynomial in $ x = (L/\xi)^{1/\nu} $.

In tables \ref{tab:FITPARAM-half} and \ref{tab:FITPARAM-quarter}, and in Fig.\ 1 in the main text, we present fits with a goodness-of-fit value $ p \geq 0.05 $. We emphasize that, for each energy, we have taken a single wave function from each sample to avoid inter-sample correlations \cite{Rodriguez2011}. For each energy value, we use the smallest concentration interval that yields the smallest uncertainties in $ n_\text{c} $ and $ \nu $. We also make sure that the estimates of the critical parameters do not change, within error bars, with larger concentration intervals (\emph{robust} fits) and when increasing the expansion order of the scaling function (\emph{stable} fits). Due to the well-know gap estimation problem in DFT, our $n_\text{c}$ values are overestimated by roughly a factor $2$ \cite{Engel2011}. In some cases for $ \SI{-0.15}{eV} \lesssim \varepsilon \lesssim \SI{-0.1}{eV}$, especially for $ \lambda = 1/4 $, the critical concentration is close to the lowest achievable doping, where $ N_\text{S} = 1 $. The few concentrations available in these cases prevent us from obtaining acceptable fits ($ p \geq 0.05 $).
Fits using scaling functions with irrelevant scaling terms \cite{Slevin1999,Rodriguez2011} converge, but do not consistently meet the stability criterion for the available system sizes.

%%%%%%%%%%%%%%%%%%%%%%%%%%%%%%%%%%%%%%%%%%%%%%%%%%%%%%%%%%%%
%%%%%%%%%%%%%%%%%%%%%%%%%%%%%%%%%%%%%%%%%%%%%%%%%%%%%%%%%%%%

\bibliographystyle{naturemag}

% %%%%%%%%%%%%%%%%%%%%%%%%%%%%%%%%%%%%%%%%%%%%%%%%%%%%%%%%%%%%

%%%%%%%%%%%%%%%%%%%%%%%%%%%%%%%%%%%%%%%%%%%%%%%%%%%%%%%%%%%%
\begin{widetext}
	\begin{table*}
		\caption{\label{tab:FITPARAM-half} Summary of the fit results for $ q = 0 $ listed by decreasing energy and coarse-graining $ \lambda $. The system sizes used are $L=16$ to 22 and the concentrations used for each energy lie in the interval $ (n_\text{min},n_\text{max}) $. Uncertainties on the critical exponent $ \nu $ and concentration $ n_\text{c} $ are 95\% confidence intervals. Energies are expressed in $ \si{eV} $ and all concentrations in units of $ \SI{E20}{cm^{-3}} $. $N_\text{P}$ and $N_\text{D}$ indicate, respectively, the number of parameters and of data points used (with average percentage precision in parenthesis), while $\chi^2$ and $p$ are the values of the $\chi^2$ statistics and the goodness-of-fit probability. The expansion is in the order $m_L$, $m_\rho$.}
		\begin{tabular}{
				@{}
				S[table-format=2.3]
				S[table-format=1.1]
				S[table-format=2.1]@{\(,\)\,}
				S[table-format=2.1]
				S[table-format=1.2]@{\,\(\pm\)\,}
				S[table-format=1.2]
				S[table-format=2.2]@{\,\(\pm\)\,}
				S[table-format=1.2]
				S[table-format=1.0]
				S[table-format=3.0]@{\,\((\)}
				S[table-format=1.2]@{\()\)\,}
				S[table-format=4.0]
				S[table-format=1.2]
				c
				@{}
			}
			\hline\hline
			{$ \varepsilon-\varepsilon_\text{F}$ } &
			$ \lambda $ &
			\multicolumn{2}{c}{$n_\text{min}$, $n_\text{max}$}&
			\multicolumn{2}{c}{$\nu$} &
			\multicolumn{2}{c}{$n_\text{c}$ } & 
			$ N_\text{P} $ &
			\multicolumn{2}{c}{$ N_\text{D} $ (prec.) } &
			$ \chi^{2} $ &
			$ p $ &
			$m_L$ $m_\rho$  \\
			\hline
			0.000  & 0.5 & 7.6 & 14.0 & 0.54 & 0.14 & 10.79 & 0.44 & 6 & 72  & 0.11 & 73  & 0.30 & 3 1 \\
			-0.011 & 0.5 & 7.7 & 14.3 & 0.58 & 0.14 & 11.01 & 0.44 & 6 & 72  & 0.11 & 62  & 0.61 & 3 1 \\
			-0.023 & 0.5 & 8.3 & 13.8 & 0.68 & 0.19 & 11.01 & 0.43 & 6 & 58  & 0.11 & 65  & 0.11 & 3 1 \\
			-0.034 & 0.5 & 7.8 & 13.0 & 0.73 & 0.18 & 10.38 & 0.37 & 6 & 62  & 0.12 & 59  & 0.36 & 3 1 \\
			-0.045 & 0.5 & 7.3 & 12.1 & 0.75 & 0.17 & 9.69  & 0.33 & 6 & 68  & 0.13 & 81  & 0.05 & 3 1 \\
			-0.057 & 0.5 & 6.8 & 10.2 & 0.74 & 0.23 & 8.60  & 0.34 & 6 & 68  & 0.13 & 80  & 0.06 & 3 1 \\
			-0.068 & 0.5 & 6.2 & 9.3  & 0.81 & 0.25 & 7.79  & 0.30 & 6 & 84  & 0.13 & 94  & 0.11 & 3 1 \\
			-0.079 & 0.5 & 5.1 & 7.7  & 0.85 & 0.33 & 6.43  & 0.26 & 6 & 90  & 0.12 & 88  & 0.35 & 3 1 \\
			-0.102 & 0.5 & 0.2 & 1.4  & 1.01 & 0.26 & 0.82  & 0.08 & 6 & 45  & 0.43 & 27  & 0.92 & 3 1 \\
			-0.113 & 0.5 & 0.2 & 1.4  & 0.77 & 0.16 & 0.82  & 0.05 & 6 & 45  & 0.55 & 39  & 0.46 & 3 1 \\
			-0.125 & 0.5 & 0.4 & 1.6  & 0.69 & 0.13 & 1.00  & 0.06 & 6 & 52  & 0.69 & 53  & 0.21 & 3 1 \\
			-0.136 & 0.5 & 0.6 & 1.8  & 0.91 & 0.23 & 1.21  & 0.08 & 6 & 50  & 0.59 & 56  & 0.10 & 3 1 \\
			-0.147 & 0.5 & 0.8 & 2.3  & 1.34 & 0.39 & 1.48  & 0.11 & 6 & 54  & 0.59 & 63  & 0.07 & 3 1 \\
			-0.159 & 0.5 & 1.1 & 2.5  & 1.26 & 0.35 & 1.84  & 0.12 & 6 & 47  & 0.46 & 48  & 0.22 & 3 1 \\
			-0.170 & 0.5 & 1.3 & 3.1  & 1.16 & 0.27 & 2.21  & 0.12 & 6 & 49  & 0.42 & 45  & 0.38 & 3 1 \\
			-0.181 & 0.5 & 1.6 & 3.6  & 1.54 & 0.40 & 2.65  & 0.16 & 6 & 53  & 0.38 & 45  & 0.57 & 3 1 \\
			-0.193 & 0.5 & 2.3 & 4.3  & 1.73 & 0.57 & 3.29  & 0.20 & 6 & 47  & 0.29 & 53  & 0.10 & 3 1 \\
			-0.204 & 0.5 & 2.3 & 4.9  & 1.54 & 0.36 & 3.65  & 0.18 & 6 & 61  & 0.32 & 71  & 0.07 & 3 1 \\
			-0.215 & 0.5 & 3.1 & 5.7  & 1.46 & 0.37 & 4.43  & 0.21 & 6 & 67  & 0.27 & 52  & 0.79 & 3 1 \\
			-0.227 & 0.5 & 3.3 & 6.8  & 1.49 & 0.28 & 5.06  & 0.20 & 6 & 102 & 0.24 & 119 & 0.06 & 3 1 \\
			-0.238 & 0.5 & 3.9 & 8.1  & 1.58 & 0.29 & 5.90  & 0.21 & 6 & 127 & 0.21 & 131 & 0.25 & 3 1 \\
			-0.249 & 0.5 & 5.0 & 8.4  & 1.25 & 0.30 & 6.72  & 0.23 & 6 & 106 & 0.20  & 100 & 0.48 & 3 1 \\
			-0.261 & 0.5 & 6.1 & 9.1  & 1.32 & 0.43 & 7.68  & 0.33 & 6 & 86  & 0.20  & 86  & 0.31 & 3 1 \\
			-0.272 & 0.5 & 5.4 & 11.2 & 1.68 & 0.33 & 8.32  & 0.38 & 6 & 131 & 0.22 & 140 & 0.17 & 3 1 \\
			-0.283 & 0.5 & 6.1 & 11.3 & 1.43 & 0.30 & 8.62  & 0.35 & 6 & 107 & 0.22 & 97  & 0.60 & 3 1 \\
			-0.295 & 0.5 & 6.6 & 11.0 & 1.33 & 0.34 & 8.83  & 0.38 & 6 & 84  & 0.24 & 69  & 0.75 & 3 1 \\
			-0.317 & 0.5 & 8.0 & 12.0 & 1.25 & 0.46 & 9.97  & 0.46 & 6 & 52  & 0.24 & 51  & 0.30 & 3 1 \\
			0.000  & 0.25 & 7.4 & 13.7 & 0.44 & 0.07 & 10.48 & 0.25 & 6 & 77  & 0.09 & 72  & 0.45 & 3 1 \\
			-0.011 & 0.25 & 7.4 & 13.8 & 0.48 & 0.07 & 10.57 & 0.22 & 6 & 74  & 0.09 & 77  & 0.21 & 3 1 \\
			-0.023 & 0.25 & 7.4 & 13.7 & 0.66 & 0.08 & 10.53 & 0.23 & 6 & 77  & 0.09 & 89  & 0.07 & 3 1 \\
			-0.034 & 0.25 & 6.9 & 12.9 & 0.65 & 0.06 & 9.98  & 0.24 & 7 & 85  & 0.09 & 96  & 0.08 & 3 2 \\
			-0.045 & 0.25 & 6.8 & 11.3 & 0.72 & 0.09 & 9.09  & 0.18 & 6 & 80  & 0.09 & 95  & 0.05 & 3 1 \\
			-0.057 & 0.25 & 6.0 & 10.0 & 0.72 & 0.09 & 7.98  & 0.14 & 6 & 99  & 0.09 & 110 & 0.10 & 3 1 \\
			-0.068 & 0.25 & 4.6 & 8.5  & 0.89 & 0.12 & 6.75  & 0.16 & 6 & 64  & 0.09 & 74  & 0.07 & 3 1 \\
			-0.159 & 0.25 & 1.7 & 4.1  & 1.20 & 0.17 & 2.89  & 0.10 & 7 & 52  & 0.14 & 61  & 0.06 & 3 2 \\
			-0.181 & 0.25 & 2.9 & 4.8  & 1.09 & 0.21 & 3.93  & 0.13 & 7 & 47  & 0.13 & 54  & 0.07 & 3 2 \\
			-0.215 & 0.25 & 4.3 & 7.1  & 0.91 & 0.12 & 5.59  & 0.11 & 7 & 94  & 0.11 & 109 & 0.05 & 4 1 \\
			-0.227 & 0.25 & 4.4 & 8.2  & 0.96 & 0.10 & 6.31  & 0.11 & 7 & 116 & 0.11 & 130 & 0.08 & 3 2 \\
			-0.238 & 0.25 & 5.2 & 8.6  & 0.88 & 0.11 & 6.86  & 0.12 & 6 & 107 & 0.12 & 110 & 0.25 & 3 1 \\
			-0.249 & 0.25 & 6.1 & 9.1  & 0.93 & 0.17 & 7.60  & 0.16 & 6 & 86  & 0.12 & 98  & 0.08 & 3 1 \\
			-0.261 & 0.25 & 6.7 & 10.1 & 1.03 & 0.19 & 8.32  & 0.20 & 6 & 70  & 0.12 & 70  & 0.28 & 3 1 \\
			-0.272 & 0.25 & 7.2 & 10.8 & 0.84 & 0.13 & 9.09  & 0.22 & 7 & 59  & 0.13 & 65  & 0.11 & 3 2 \\
			-0.283 & 0.25 & 7.5 & 11.3 & 0.90 & 0.15 & 9.36  & 0.20 & 6 & 55  & 0.14 & 63  & 0.08 & 3 1 \\
			-0.295 & 0.25 & 7.7 & 11.5 & 0.96 & 0.16 & 9.58  & 0.21 & 6 & 54  & 0.14 & 39  & 0.81 & 3 1 \\
			-0.306 & 0.25 & 8.2 & 12.2 & 0.90 & 0.17 & 10.19 & 0.22 & 6 & 49  & 0.14 & 38  & 0.69 & 3 1 \\
			-0.317 & 0.25 & 8.0 & 13.4 & 1.07 & 0.17 & 10.72 & 0.26 & 6 & 60  & 0.14 & 59  & 0.28 & 3 1 \\
			-0.329 & 0.25 & 9.0 & 13.4 & 1.10 & 0.24 & 11.17 & 0.30 & 6 & 45  & 0.14 & 33  & 0.74 & 3 1 \\
			\hline\hline
		\end{tabular}
		
	\end{table*} 
\end{widetext}
%%%%%%%%%%%%%%%%%%%%%%%%%%%%%%%%%%%%%%%%%%%%%%%%%%%%%%%%%%%%

%%%%%%%%%%%%%%%%%%%%%%%%%%%%%%%%%%%%%%%%%%%%%%%%%%%%%%%%%%%%
\begin{widetext}
	\begin{table*}
		\caption{\label{tab:FITPARAM-quarter}  Summary of the fit results for $ q = 1 $ listed by decreasing energy and coarse-graining $ \lambda $. The table is formatted as in Tab.\ \ref{tab:FITPARAM-half}.}
		\begin{tabular}{
				@{}
				S[table-format=2.3]
				S[table-format=1.1]
				S[table-format=2.1]@{\(,\)\,}
				S[table-format=2.1]
				S[table-format=1.2]@{\,\(\pm\)\,}
				S[table-format=1.2]
				S[table-format=2.2]@{\,\(\pm\)\,}
				S[table-format=1.2]
				S[table-format=1.0]
				S[table-format=3.0]@{\,\((\)}
				S[table-format=1.2]@{\()\)\,}
				S[table-format=4.0]
				S[table-format=1.2]
				c
				@{}
			}
			\hline\hline
			{$ \varepsilon-\varepsilon_\text{F}$ } &
			$ \lambda $ &
			\multicolumn{2}{c}{$n_\text{min}$, $n_\text{max}$}&
			\multicolumn{2}{c}{$\nu$} &
			\multicolumn{2}{c}{$n_\text{c}$ } & 
			$ N_\text{P} $ &
			\multicolumn{2}{c}{$ N_\text{D} $ (prec.)} &
			$ \chi^{2} $ &
			$ p $ &
			$m_L$ $m_\rho$  \\
			\hline
			%\endlastfoot \\ \\
			0.000  & 0.5  & 7.5 & 13.9 & 0.55 & 0.14 & 10.72 & 0.45 & 6 & 76  & 0.12 & 78  & 0.24 & 3 1 \\
			-0.011 & 0.5  & 8.1 & 13.5 & 0.49 & 0.13 & 10.82 & 0.39 & 6 & 60  & 0.12 & 51  & 0.61 & 3 1 \\
			-0.023 & 0.5  & 8.8 & 13.2 & 0.57 & 0.22 & 10.99 & 0.44 & 6 & 44  & 0.12 & 48  & 0.13 & 3 1 \\
			-0.034 & 0.5  & 8.3 & 12.4 & 0.54 & 0.18 & 10.34 & 0.35 & 6 & 48  & 0.13 & 43  & 0.43 & 3 1 \\
			-0.045 & 0.5  & 7.7 & 11.5 & 0.80 & 0.28 & 9.61  & 0.38 & 6 & 54  & 0.15 & 60  & 0.12 & 3 1 \\
			-0.057 & 0.5  & 6.8 & 10.2 & 0.71 & 0.23 & 8.56  & 0.34 & 6 & 68  & 0.15 & 80  & 0.06 & 3 1 \\
			-0.068 & 0.5  & 5.5 & 10.1 & 0.72 & 0.13 & 7.78  & 0.24 & 6 & 117 & 0.15 & 129 & 0.11 & 3 1 \\
			-0.079 & 0.5  & 4.6 & 8.5  & 0.98 & 0.26 & 6.44  & 0.27 & 6 & 119 & 0.14 & 136 & 0.07 & 3 1 \\
			-0.091 & 0.5  & 0.0 & 1.0  & 1.13 & 0.79 & 0.48  & 0.14 & 6 & 27  & 0.62 & 32  & 0.06 & 3 1 \\
			-0.102 & 0.5  & 0.1 & 1.3  & 0.89 & 0.30 & 0.68  & 0.08 & 6 & 37  & 0.67 & 18  & 0.97 & 3 1 \\
			-0.113 & 0.5  & 0.1 & 1.3  & 0.74 & 0.22 & 0.73  & 0.07 & 6 & 37  & 0.96 & 34  & 0.31 & 3 1 \\
			-0.125 & 0.5  & 0.4 & 1.4  & 0.76 & 0.22 & 0.89  & 0.08 & 6 & 45  & 1.21 & 40  & 0.44 & 3 1 \\
			-0.136 & 0.5  & 0.6 & 1.7  & 0.85 & 0.28 & 1.11  & 0.09 & 6 & 48  & 1.09 & 50  & 0.19 & 3 1 \\
			-0.147 & 0.5  & 0.7 & 2.0  & 1.49 & 0.54 & 1.36  & 0.13 & 6 & 51  & 1.15 & 55  & 0.15 & 3 1 \\
			-0.159 & 0.5  & 0.9 & 2.6  & 1.23 & 0.32 & 1.69  & 0.11 & 6 & 56  & 0.87 & 60  & 0.16 & 3 1 \\
			-0.170 & 0.5  & 1.1 & 3.2  & 1.22 & 0.27 & 2.10  & 0.12 & 6 & 62  & 0.79 & 63  & 0.24 & 3 1 \\
			-0.181 & 0.5  & 1.3 & 3.8  & 1.48 & 0.32 & 2.46  & 0.14 & 6 & 68  & 0.71 & 74  & 0.14 & 3 1 \\
			-0.193 & 0.5  & 1.6 & 4.7  & 1.49 & 0.28 & 3.07  & 0.16 & 6 & 75  & 0.54 & 88  & 0.06 & 3 1 \\
			-0.204 & 0.5  & 2.1 & 4.9  & 1.59 & 0.37 & 3.52  & 0.19 & 6 & 66  & 0.45 & 78  & 0.06 & 3 1 \\
			-0.215 & 0.5  & 3.0 & 5.6  & 1.58 & 0.48 & 4.32  & 0.23 & 6 & 64  & 0.34 & 57  & 0.52 & 3 1 \\
			-0.227 & 0.5  & 3.5 & 6.5  & 1.55 & 0.40 & 4.98  & 0.23 & 6 & 87  & 0.29 & 97  & 0.10 & 3 1 \\
			-0.238 & 0.5  & 4.1 & 7.5  & 1.72 & 0.39 & 5.81  & 0.24 & 6 & 112 & 0.26 & 106 & 0.48 & 3 1 \\
			-0.249 & 0.5  & 5.0 & 8.4  & 1.28 & 0.33 & 6.70  & 0.25 & 6 & 106 & 0.24 & 96  & 0.59 & 3 1 \\
			-0.261 & 0.5  & 6.2 & 9.2  & 1.30 & 0.45 & 7.66  & 0.33 & 6 & 83  & 0.23 & 86  & 0.22 & 3 1 \\
			-0.272 & 0.5  & 6.3 & 10.5 & 1.58 & 0.53 & 8.43  & 0.47 & 6 & 91  & 0.24 & 74  & 0.81 & 3 1 \\
			-0.283 & 0.5  & 6.1 & 11.3 & 1.52 & 0.35 & 8.61  & 0.40 & 6 & 107 & 0.27 & 104 & 0.41 & 3 1 \\
			-0.295 & 0.5  & 6.6 & 11.0 & 1.39 & 0.40 & 8.81  & 0.42 & 6 & 84  & 0.29 & 73  & 0.63 & 3 1 \\
			-0.317 & 0.5  & 8.0 & 12.0 & 1.24 & 0.49 & 9.95  & 0.49 & 6 & 52  & 0.28 & 50  & 0.32 & 3 1 \\
			0.000  & 0.25 & 7.2 & 13.4 & 0.47 & 0.08 & 10.27 & 0.27 & 6 & 79  & 0.12 & 78  & 0.32 & 3 1 \\
			-0.011 & 0.25 & 7.4 & 13.7 & 0.51 & 0.07 & 10.45 & 0.23 & 6 & 77  & 0.12 & 77  & 0.29 & 3 1 \\
			-0.023 & 0.25 & 7.4 & 13.7 & 0.69 & 0.09 & 10.49 & 0.25 & 6 & 77  & 0.13 & 80  & 0.23 & 3 1 \\
			-0.034 & 0.25 & 6.9 & 12.7 & 0.68 & 0.08 & 9.70  & 0.19 & 6 & 88  & 0.13 & 102 & 0.07 & 3 1 \\
			-0.045 & 0.25 & 7.7 & 10.4 & 0.72 & 0.24 & 9.14  & 0.25 & 6 & 42  & 0.13 & 40  & 0.29 & 3 1 \\
			-0.057 & 0.25 & 6.4 & 9.6  & 0.66 & 0.11 & 7.98  & 0.15 & 6 & 79  & 0.13 & 72  & 0.49 & 3 1 \\
			-0.068 & 0.25 & 5.4 & 8    & 0.82 & 0.18 & 6.74  & 0.15 & 6 & 92  & 0.13 & 96  & 0.21 & 3 1 \\
			-0.079 & 0.25 & 2.9 & 6.1  & 1.55 & 0.33 & 4.50  & 0.21 & 6 & 83  & 0.15 & 92  & 0.12 & 3 1 \\
			-0.102 & 0.25 & 0.5 & 1.9  & 1.24 & 0.22 & 1.09  & 0.08 & 7 & 57  & 0.32 & 54  & 0.33 & 4 1 \\
			-0.125 & 0.25 & 0.4 & 2.0  & 0.99 & 0.13 & 1.21  & 0.06 & 6 & 63  & 0.53 & 72  & 0.09 & 3 1 \\
			-0.136 & 0.25 & 0.6 & 2.2  & 1.15 & 0.21 & 1.40  & 0.07 & 6 & 61  & 0.56 & 64  & 0.20 & 3 1 \\
			-0.147 & 0.25 & 1.2 & 2.2  & 1.48 & 0.65 & 1.74  & 0.13 & 6 & 34  & 0.37 & 37  & 0.13 & 3 1 \\
			-0.170 & 0.25 & 1.6 & 3.4  & 1.40 & 0.31 & 2.52  & 0.13 & 6 & 44  & 0.32 & 39  & 0.43 & 3 1 \\
			-0.181 & 0.25 & 2.0 & 4.1  & 1.65 & 0.39 & 3.00  & 0.16 & 6 & 51  & 0.29 & 46  & 0.45 & 3 1 \\
			-0.193 & 0.25 & 2.5 & 5.1  & 1.37 & 0.21 & 3.86  & 0.13 & 7 & 63  & 0.24 & 70  & 0.10 & 3 2 \\
			-0.204 & 0.25 & 3.1 & 5.7  & 1.11 & 0.17 & 4.29  & 0.12 & 6 & 67  & 0.22 & 77  & 0.08 & 3 1 \\
			-0.215 & 0.25 & 3.5 & 6.5  & 1.22 & 0.19 & 4.94  & 0.14 & 6 & 87  & 0.20  & 73  & 0.73 & 3 1 \\
			-0.227 & 0.25 & 4.5 & 6.7  & 1.14 & 0.28 & 5.73  & 0.16 & 6 & 74  & 0.19 & 81  & 0.13 & 3 1 \\
			-0.238 & 0.25 & 4.8 & 8.0  & 1.02 & 0.15 & 6.33  & 0.13 & 6 & 104 & 0.19 & 95  & 0.57 & 3 1 \\
			-0.249 & 0.25 & 5.7 & 8.5  & 1.03 & 0.22 & 7.09  & 0.17 & 6 & 92  & 0.19 & 86  & 0.49 & 3 1 \\
			-0.261 & 0.25 & 5.9 & 9.9  & 1.17 & 0.20 & 7.85  & 0.21 & 6 & 100 & 0.20  & 94  & 0.47 & 3 1 \\
			-0.272 & 0.25 & 6.4 & 10.6 & 1.25 & 0.22 & 8.48  & 0.24 & 6 & 90  & 0.21 & 91  & 0.28 & 3 1 \\
			-0.283 & 0.25 & 6.7 & 11.1 & 1.20 & 0.19 & 8.89  & 0.23 & 6 & 83  & 0.22 & 85  & 0.25 & 3 1 \\
			-0.295 & 0.25 & 6.8 & 11.4 & 1.13 & 0.18 & 9.04  & 0.24 & 6 & 78  & 0.23 & 51  & 0.97 & 3 1 \\
			-0.306 & 0.25 & 7.3 & 12.1 & 1.22 & 0.20 & 9.68  & 0.26 & 6 & 68  & 0.24 & 65  & 0.38 & 3 1 \\
			-0.317 & 0.25 & 8.2 & 12.4 & 1.11 & 0.30 & 10.09 & 0.29 & 6 & 47  & 0.23 & 50  & 0.15 & 3 1 \\
			\hline\hline
		\end{tabular}
	\end{table*} 
\end{widetext}
%%%%%%%%%%%%%%%%%%%%%%%%%%%%%%%%%%%%%%%%%%%%%%%%%%%%%%%%%%%%

%%%%%%%%%%%%%%%%%%%%%%%%%%%%%%%%%%%%%%%%%%%%%%%%%%%%%%%%%%%%

\end{document}